**Terahertz oscillation of 180° domain walls in ferroelectric membranes**

Xiangwei Guo[1a)], Jiaxuan Wu[1a)], Yujie Zhu[1], Aiden Ross[2], Bo Wang[2], Paul G. Evans[1], Long-Qing Chen[2], and Jia-Mian Hu[1*]

*[1]Department of Materials Science and Engineering, University of Wisconsin-Madison, Madison, WI, 53706, USA*

*[2]Department of Materials Science and Engineering, The Pennsylvania State University, University Park, Pennsylvania, 16802, USA*

**Abstract**

A fundamentally intriguing yet not well understood topic in the field of ferroelectrics is the collective excitation of domain walls (DWs), with potential applications to DW-based nanoelectronic and optoelectronic devices. Here we use dynamical phase-field simulations to identify the collective modes of an Ising-type 180° DW in a uniaxially strained $BaTiO_3$ membrane. The membrane, which concurrently functions as a cavity for polarization and acoustic waves, permits cavity-enhanced resonant excitation of polarization waves. The simulation reveals an unconventional DW sliding mode that exhibits a bulk-polarization-charge-driven nonzero resonant frequency in the sub-terahertz (THz) regime with a dynamically changing internal structure during sliding. These features differ from the previously reported zero-frequency DW sliding mode or the surface-polarization-charge-driven DW sliding mode. The effect of strain on the frequency of this unconventional DW sliding mode and other previously known THz DW eigenmodes is investigated by dynamical phase-field simulations and interpreted by eigenmode analysis or analytical calculation in a simplified one-dimensional (1D) system. These results provide new insights into the high-frequency dynamics of ferroelectric DWs and suggest opportunities for realizing strain control of phonon-DW resonance, and more broadly, for discovering and controlling unconventional DW modes in conventional domain patterns with applications to reconfigurable THz and optical devices.

[*]E-mail: jhu238@wisc.edu

[a)]X. Guo and J. Wu contribute equally to this work.

## 1. Introduction

Ferroic membranes can be integrated with materials with different crystal structure and composition through lift-off and transferring [1-6] and can enable large, continuously tunable strain, strain gradient and more complex deformation modes [7] (e.g., stretching [8-10], bending [11,12], rippling [13,14], folding [15], and twisting [16]) that are difficult to achieve in films grown on an elastically stiff substrate [17]. Capitalizing on these capabilities, ferroic membranes have thus far been exploited for stabilizing topologically nontrivial structures [5,16,18] and realizing extreme-state strain engineering of materials properties [15,19-21], with potential applications to flexible electronic, spintronic, and micro-electromechanical devices [22-27]. The high-frequency dynamics of these membranes may exhibit novel time-dependent phenomena not available in supported ferroic thin films but are comparatively unexplored.

Dynamical phase-field modeling (DPFM) has been extensively used to predict the responses of polarization (especially the topological polar textures [28-32]) to external stimuli in ferroelectric bulk crystals [33-35] and thin films [36] as well as in strained ferroelectric/dielectric superlattice thin films [28-30,32,37]. Recently, by incorporating the coupled dynamics of polarization, strain, and electromagnetic (EM) waves [38], DPFM has been extended to enable the study of EM wave transmission through a ferroelectric material [39-41]. Molecular dynamics simulations have also been used to predict the formation and switching dynamics of ferroelectric domains in bulk crystals [42-45], strained thin films [46-50] and ferroelectric/dielectric superlattice thin films [51,52]. Experimentally, new collective modes have recently been discovered [28,31,37] through the excitation of topological polar textures (vortices and skyrmions) in $PbTiO_3/SrTiO_3$ superlattice thin films by a THz electric-field pulse. However, there have been few studies of the coupled dynamics of polarization waves and acoustic waves in multidomain ferroelectric membranes.

Here, we employ DPFM to simulate the coupled polarization-strain dynamics in the terahertz (THz) regime in uniaxially strained $BaTiO_3$ membranes with Ising-type 180º domain walls (DWs). We evaluate the frequency-wavevector dispersion relation of the polarization waves from the simulation results. The modes of polarization waves include both the intrinsic modes, i.e., those that would appear in a single-domain ferroelectric, and the waves confined at the DWs. Guided by the dispersion relation, we further simulate the excitation of individual DW modes by a broad-band THz pulse and reconstruct the mode profiles. These modes include the conventional THz eigenmodes of a 180º DW that can be predicted by eigenmode analysis [53], and notably, an unconventional sub-THz DW sliding mode. The frequencies of these DW modes can be effectively tuned by strain, which is predicted by dynamical phase-field simulations and interpreted by simplified eigenmode analysis and an analytical model. Our results provide new insights into the collective excitation of ferroelectric DWs and suggest potential applications of domain-engineered ferroelectric nanomembranes in wave-based computing [54] and THz wave generation [55].

## 2. Dynamical phase-field simulations

In displacive ferroelectrics, the lattice polarization $P_i$ is proportional to the coordinates of the soft mode $Q_j$ (unit: m) via $P_i$=$Z_{ij}Q_j$, where $Z_{ij}$=$e_c Z^*_{ij}/V$ is the reduced Born effective charge tensor (typically diagonal) and has a unit of C/m$^3$, describing the charge density associated with the soft mode [40]. $Z^*_{ij}$ is the dimensionless Born effective charge, $e_c$ is the elementary charge, and V is the volume of the unit cell. The subscripts $i$,$j$=1,2,3 denote the three orthogonal axes ($x_1$-$x_2$-$x_3$) in the crystal physical systems of the parent paraelectric phase. In a dynamical phase-field model, the ferroelectric domain structure is represented by the spatial distribution of $P_i$ in each discretized cell of the simulation system. The temporal evolution of $P_i$ is governed by [35,40],

$$\mu_{ij}\frac{\partial^2 P_j}{\partial t^2}+\gamma_{ij}\frac{\partial P_j}{\partial t}=-\frac{\delta F_{\text{tot}}}{\delta P_i}. \quad (1)$$

Here, $\mu_{ij}$=$M_\text{p}$/($Z_{ij}^2V$) is the mass coefficient tensor [56], where $M_\text{p}$ is the effective mass of the soft mode (unit: kg), and $V$ is the volume of the unit cell. $\gamma_{ij}$ is the phenomenological damping coefficient tensor. $\gamma_{ij}$ characterizes the energy dissipation rate from the soft-mode polarization and can be determined based on the linewidth of the soft-mode phonon in the dielectric or THz transmission spectra of a single-domain bulk ferroelectric crystal. A relevant example is in the case of $LiTaO_3$ single crystals [57]. For simplicity, here we consider both $\mu_{ij}(\equiv\mu)$ and $\gamma_{ij}$ $(\equiv\gamma)$ to be scalars, as assumed in existing theoretical and computational works (e.g., [35,38,39]).

$F_{\text{tot}}$ is the total Helmholtz free energy of the ferroelectric, which has temperature $T$, $P_i$, electric field $E_i$, and total strain $\varepsilon_{ij}$ as the natural variables. Using the Gibbs free energy of the ferroelectric under zero $P_i$, zero electric field, and zero stress, as the reference, $F_{\text{tot}}$ can be expressed as $F_{\text{tot}}=\int\left(f^{\text{E}}+f_{\text{grad}}\right)dV$, where $f^{\text{E}}$=$f_{\text{Landau}}+f_{\text{elec}}+f_{\text{elas}}$ is the Helmholtz free energy density [39] in a local volume with spatially uniform physical quantities. $f_{\text{Landau}}$ is the Landau free energy density. For $BaTiO_3$, the expression of $f_{\text{Landau}}$ is provided in [39]. The electrostatic energy density is given by $f_{\text{elec}}$=$-\frac{1}{2}\kappa_0\kappa_\text{b}E_iE_j-E_iP_i$, where the background dielectric permittivity $\kappa_\text{b}$ describes the contribution from the electronic polarization and polarization arising from lattice phonon modes other than soft mode. The electric field $E_i$ is the sum of the externally applied electric field $E_i^{\text{inc}}$, the depolarization field $E_i^{\text{d}}$, and the radiating electric field $E_i^{\text{rad}}$. Here, the local depolarization field $E_i^{\text{d}}$ is obtained by numerically solving the electrostatic equilibrium equation $\nabla\cdot\left(\kappa_0\kappa_\text{b}E_i^{\text{d}}+P_i(\mathbf{r})\right)=0$, where the effects of long-range electrostatic interaction among the spatially inhomogeneous electric polarization $P_i(\mathbf{r})$ are incorporated. $E_i^{\text{rad}}$ is generated by dynamical oscillation of polarization (the polarization current) and can be evaluated by solving the complete set of Maxwell's equations [38,41]. The elastic energy density is expressed as $f_{\text{elast}}=\frac{1}{2}c_{ijkl}(\varepsilon_{kl}-\varepsilon_{kl}^0)(\varepsilon_{ij}-\varepsilon_{ij}^0)$, where the local total strain $\varepsilon_{ij}$ describes the deformation with respect to the parent paraelectric phase, and $\varepsilon_{ij}^0=Q_{ijkl}P_kP_l$ is the eigenstrain describing the mechanical deformation arising due to the paraelectric-to-ferroelectric phase transition under zero stress.

In the framework of Landau-Ginzburg-Devonshire (LGD) theory, $c_{ijkl}$ is the elastic stiffness tensor under zero spontaneous polarization [58,59] and hence should have the symmetry of the parent paraelectric phase (which is cubic for $BaTiO_3$). The electrostrictive tensor $Q_{ijkl}$ is also a quantity determined under zero spontaneous polarization and therefore has a cubic symmetry. In this work, the elastic stiffness tensor in the tetragonal and other ferroelectric phases (orthorhombic and rhombohedral) in thermodynamic equilibrium is the constant electric-field elastic stiffness tensor [58], defined as $c_{ijkl}^{\text{E}}=\left(\frac{\partial^2 f^{\text{E}}}{\partial\varepsilon_{kl}\partial\varepsilon_{ij}}\right)_{T,P_i,E_i}$. $c_{ijkl}^{\text{E}}$ is a function of both $Q_{ijkl}$ and $P_i$, and has the symmetry of the specific ferroelectric phase. The gradient energy density is given by $f_{\text{grad}}=\frac{1}{2}G_{ijkl}\frac{\partial P_i}{\partial x_j}\frac{\partial P_k}{\partial x_l}$ [60], where $G_{ijkl}$ is the gradient energy coefficient tensor determined under zero spontaneous polarization. For ferroelectrics with a cubic parent paraelectric phase, $f_{\text{grad}}$ can be expanded into,

$$f_{\mathrm{grad}} = \frac{1}{2} G_{ijkl} P_{i,j} P_{k,l} = \frac{1}{2} G_{11} \left( P_{1,1}^2 + P_{2,2}^2 + P_{3,3}^2 \right) + G_{12} \left( P_{1,1} P_{2,2} + P_{2,2} P_{3,3} + P_{1,1} P_{3,3} \right)$$
$$+ \frac{1}{2} G_{44} \left[ \left( P_{1,2} + P_{2,1} \right)^2 + \left( P_{2,3} + P_{3,2} \right)^2 + \left( P_{1,3} + P_{3,1} \right)^2 \right], \quad (2)$$

where $P_{i,j} = \partial P_i / \partial x_j$. In the present dynamical phase-field simulations, we assume an isotropic gradient energy density (similar to the treatment in [61]), i.e., $f_{\mathrm{grad}}=\frac{1}{2} G_0 (\nabla \mathbf{P})^2$, where $G_0$ is the isotropic gradient energy coefficient.

To simulate the coupled polarization-strain dynamics in ferroelectric membranes, the dynamical total strain $\Delta\varepsilon_{ij}(t)$ needs to be considered in calculating $f_{\mathrm{elas}}$. We begin by writing the time-varying local total strain as $\varepsilon_{ij}(t) = \varepsilon_{ij}(t=0) + \Delta\varepsilon_{ij}(t)$. In simulating the ferroelectric domain pattern formation during a paraelectric-to-ferroelectric phase transformation, $\varepsilon_{ij}(t=0)$ is the local total strain at the initial state ($t$=0), which is a paraelectric phase with a small, randomly fluctuating polarization magnitude $|P_i(\mathbf{r}, t=0)| \leq 0.01$ C/m$^2$. The resulting eigenstrain in the initial state, $\varepsilon_{ij}^0(t=0)$, is therefore small. As a result, $\varepsilon_{ij}(t=0)$ should be spatially uniform and can be calculated directly based on the mechanical boundary condition. Specifically, for a stress-free ferroelectric membrane, one has $\varepsilon_{ij}(t=0)=\varepsilon_{ij}^0$ in every cell of the simulation system, which is approximately zero because $|P_i(\mathbf{r}, t=0)|$ is close to 0 herein. For a uniaxially strained membrane where the strain is applied along the $[100]_c$ axis, the surface tractions along both the $x_2$ and $x_3$ axis are zero, i.e., $\sigma_{i2} = \sigma_{i3} = 0$ ($i$=1,2,3) [39], and one has,

$$\varepsilon_{11}(t=0) = \varepsilon_{11}^{\mathrm{tot}}, \quad (3a)$$

$$\varepsilon_{12}(t=0) = \varepsilon_{13}(t=0) = \varepsilon_{23}(t=0) = 0, \quad (3b)$$

$$\varepsilon_{22}(t=0) = \varepsilon_{33}(t=0) = -\frac{c_{12}}{c_{11}+c_{12}} \varepsilon_{11}^{\mathrm{tot}}, \quad (3c)$$

A knowledge of $\varepsilon_{ij}(t=0)$ allows us to calculate the initial displacement $\varepsilon_{ij}(t=0)$ and the initial stress $\sigma_{ij}(t=0)$. In our dynamical phase-field simulations shown in the main text, $\varepsilon_{11}^{\mathrm{tot}}$ varies from 0.5% to 1.1% (these values are not rounded). The dynamical strain $\Delta\varepsilon_{ij}(t)$, which results from the evolving polarization $P_i(\mathbf{r}, t)$, can be obtained by solving the elastodynamic equation,

$$\rho \frac{\partial^2 \Delta u_i}{\partial t^2} = \nabla \cdot \left( \Delta\sigma_{ij} + \beta \frac{\partial \Delta\sigma_{ij}}{\partial t} \right), \quad (4)$$

where $\rho$ is the mass density and $\beta$ is the elastic damping coefficient. $\beta$ characterizes the energy dissipation rate of the acoustic phonon and can be determined based on the linewidth of the acoustic phonon in the Brillouin light scattering spectrum of a single-domain bulk ferroelectric crystal [62]. $\Delta u_i = u_i - u_i(t=0)$ is the dynamical displacement, and $\Delta\sigma_{ij} = \sigma_{ij} - \sigma_{ij}(t=0)$ is the dynamical stress and can be also written as $\Delta\sigma_{ij} = c_{ijkl}(\Delta\varepsilon_{kl} - \Delta\varepsilon_{kl}^0)$, with $\Delta\varepsilon_{kl}^0 = \varepsilon_{kl}^0 - \varepsilon_{kl}^0(t=0)$ and $\Delta\varepsilon_{kl} = \frac{1}{2}\left(\frac{\partial \Delta u_k}{\partial x_l} + \frac{\partial \Delta u_l}{\partial x_k}\right)$. Equations (1) and (4) are solved in a coupled fashion until the formation of equilibrium ferroelectric domain patterns, where the polarization-changing rate $\frac{\partial P_i}{\partial t}$, velocity field $\frac{\partial u_i}{\partial t}$ and stress-changing rate $\frac{\partial \sigma_{ij}}{\partial t}$ should all decrease to almost zero. The spatial distributions of $P_i(\mathbf{r})$, $u_i(\mathbf{r})$, and $\sigma_{ij}(\mathbf{r})$ corresponding to the equilibrium ferroelectric domain can then be used as the initial condition for the new simulations, for example, the excitation of the equilibrated ferroelectric domains by an applied electric-field pulse or other external stimuli (e.g., temperature and stress fields).

In 3D simulations, discretized cells of $N_1\Delta x_1 \times N_2\Delta x_2 \times N_3\Delta x_3$, with a cell size of $\Delta x_1 = \Delta x_3 =$ 2 nm, $\Delta x_2$ =0.4 nm, and ($N_1$, $N_2$, $N_3$=64, 320, 16), are used to describe the ferroelectric membrane. Additional vacuum layers, each having a thickness of 5 cells, are added around the ferroelectric membrane except those along the axis of the applied strain (i.e., the $x_1$ axis). The electrostatic equilibrium equation ($\nabla \cdot D_i$=0) is numerically solved by the Fourier Spectral Iterative Perturbation method [63], which requires the application of periodic boundary conditions to all the three orthogonal directions of the system. A large $\kappa_\mathrm{b}$ of >$10^5$ is assigned to the vacuum layers while we set $\kappa_\mathrm{b}$=5 for the ferroelectric BTO membrane [41]. As a result, the polarization charges at the $x_1$-$x_2$ and $x_1$-$x_3$ surfaces are fully compensated, which in practice could occur by the accumulation of mobile ionic/electronic defects at the surfaces. The equation of motion for the polarization (Eq. (1)) and the dynamical displacement (Eq. (4)) are both solved in real-space, where a classical Runge-Kutta method is used for the time marching using a time step of $\Delta t$=1.0×$10^{-15}$ s, and the central finite-difference is used to calculate the spatial derivatives in the calculation of the thermodynamic driving force resulting from $f_\mathrm{grad}$. Our control simulations show that halving $\Delta t$ does not alter the equilibrium ferroelectric domain structure. The Neumann boundary condition $\partial P_i/\partial \mathbf{n} = 0$ [41] is applied to the membrane/air interface in solving Eq. (1). Stress continuity condition, $\boldsymbol{\sigma}^{\boldsymbol{A}} \cdot \mathbf{n} = \boldsymbol{\sigma}^{\boldsymbol{B}} \cdot \mathbf{n}$, is applied to all membrane/vacuum interfaces in solving Eq. (4), where the superscripts 'A' and 'B' denote different materials and $\mathbf{n}$ is the interface normal. By using a negligibly small $c_{ijkl}$ (~ 1 Pa) for the surrounding vacuum layer, the mechanical boundary condition of zero surface traction along both the $x_2$ and $x_3$ axis are automatically satisfied. In simulating the equilibrium domain pattern [Figs. 1(a) and 2(a)], the total number of numerical time steps is set to 60,000, which is sufficiently large to ensure the system has reached the thermodynamic equilibrium. All the real-space and Fourier-space numerical solvers are accelerated through graphics processing unit (GPU) parallelization.

The material parameters for performing dynamical phase-field simulations of ferroelectric BTO are summarized in [35,38,64,65]. The calculated free energy density profile [Fig. 1(b)] is a sum of $f_\mathrm{Landau}$ and $f_\mathrm{elas}$ by setting $P_2$=$P_3$=0, a spatially uniform $P_1$, and a fixed total strain $\varepsilon_{ij}$=$\varepsilon_{ij}^0(P = P_i^\mathrm{eq})$, where $P_i^\mathrm{eq}$ is the equilibrium polarization that can be obtained by minimizing $f^\mathrm{E}$ under the electric and mechanical boundary conditions described above. For $BaTiO_3$, the Landau parameters in $f_\mathrm{Landau}$ are [65]: $\alpha_1$=4.124×$10^5$ ($T$-388) N·m$^2$·C$^{-2}$, $\alpha_{11}$=-2.097×$10^8$ N·m$^6$·C$^{-4}$, $\alpha_{12}$ =7.974×$10^8$ N·m$^6$·C$^{-4}$, $\alpha_{111}$ =1.294×$10^9$ N·m$^{10}$·C$^{-6}$, $\alpha_{112}$ =-1.950×$10^9$ N·m$^{10}$·C$^{-6}$, $\alpha_{123}$ =-2.500×$10^9$ N·m$^{10}$·C$^{-6}$, $\alpha_{1111}$ =3.863×$10^{10}$ N·m$^{14}$·C$^{-8}$, $\alpha_{1112}$ =2.529×$10^{10}$ N·m$^{14}$·C$^{-8}$, $\alpha_{1122}$=1.637×$10^{10}$ N·m$^{14}$·C$^{-8}$, $\alpha_{1123}$=1.367×$10^{10}$ N·m$^{14}$·C$^{-8}$. The elastic stiffness coefficient and the electrostrictive coefficients are (in Voigt notation) [64]: $c_{11}$= 178 GPa, $c_{12}$=96.4 GPa, $c_{44}$=122 GPa, $Q_{11}$ =0.1 m$^4$·C$^{-2}$, $Q_{12}$ =-0.034 m$^4$·C$^{-2}$, $Q_{44}$ =0.029 m$^4$·C$^{-2}$. The isotropic gradient energy coefficient is $G_0$=3.0×$10^{-11}$ C$^{-2}$·m$^4$·N, which is determined based on the comparison to inelastic neutron scattering measurement (see details in Sec. 3.1). The polarization damping coefficient $\gamma$ is 2×$10^{-7}$ Ω·m [38], which is set to the same as the reported value for $PbTiO_3$ [28] as an approximation. The elastic damping coefficient $\beta$ is 6×$10^{-12}$ s [35].

## 3. Analytical Models

### 3.1. Analytical calculation of the dispersion relation of transverse and longitudinal polarization waves in single-domain ferroelectrics

Let us consider a BTO nanomembrane with a spatially uniform initial equilibrium polarization aligning along the $x_1$ axis ($P_1^{\mathrm{eq}}$), as shown in Fig. 1(a). Our analytical model is based on a one-dimensional (1D) system with spatially uniform physical quantities along the $x_1$ and $x_3$ axes. The transverse polarization wave, $\Delta P_1(x_2,t)$, corresponds to the ferroelectric soft mode (a transverse optical phonon [66], denoted as the $A_1$(TO) mode). The longitudinal polarization wave, $\Delta P_2(x_2,t)$, is denoted as the E(LO) mode. Assuming $\gamma$=0, $E_i^{\mathrm{inc}}$=0, $E_i^{\mathrm{d}}$=0, and $E_i^{\mathrm{rad}}$=0, performing Taylor expansion for the functional derivative on the right of Eq. (1), and keeping only the linear terms, Eq. (1) reduces to,

$$\mu\frac{\partial^2 P_i}{\partial t^2}+\gamma\frac{\partial P_i}{\partial t}=G_{11}\frac{\partial^2 P_i}{\partial x_i^2}+G_{12}\frac{\partial}{\partial x_i}\sum_{j\neq i}\frac{\partial P_j}{\partial x_j}+G_{44}\sum_{i\neq j}\frac{\partial}{\partial x_j}\left(\frac{\partial P_j}{\partial x_i}+\frac{\partial P_i}{\partial x_j}\right)-\sum_j K_{ij}\Delta P_j\ \ldots, \tag{5}$$

where $K_{ij}=\partial^2 f^{\mathrm{E}}/(\partial P_i \partial P_j)$. In the 1D system herein, Eq. (5) simplifies to,

$$\mu\frac{\partial^2 P_1}{\partial t^2}+\gamma\frac{\partial P_1}{\partial t}=G_{44}\frac{\partial^2 P_1}{\partial x_2^2}-(K_{11}\Delta P_1+K_{12}\Delta P_2+K_{13}\Delta P_3), \tag{6a}$$

$$\mu\frac{\partial^2 P_2}{\partial t^2}+\gamma\frac{\partial P_2}{\partial t}=G_{11}\frac{\partial^2 P_2}{\partial x_2^2}-(K_{21}\Delta P_1+K_{22}\Delta P_2+K_{23}\Delta P_3). \tag{6b}$$

The lattice polarization, $P_i$, which represents the volumetric density of electric dipoles in each unit cell, is a discretized quantity. Therefore, we denote $P_i(x_2=na)\equiv P_{i,n}$, where $a$ is the finite spatial element along the $x_2$ direction and $n\in\mathbb{Z}$ is an integer. We then use the central finite-difference method to evaluate the second-order spatial derivatives in Eqs. (6a-b).

The plane-wave solutions to Eqs. (6a-b) are given by $P_{i,\mathrm{n}}(t)=P_i^{\mathrm{eq}}+\Delta P_{i,\mathrm{n}}=P_i^{\mathrm{eq}}+\Delta P_i^0 e^{\mathbf{i}(k_2 na-\omega t)}$, with $\omega\equiv\omega_t$ for $i$=1,3 and $\omega\equiv\omega_l$ for $i$=2. $k_2\in\mathbb{R}$ is the real wave number along the $x_2$ direction. Pugging in these plane-wave solutions and considering that the off-diagonal components of the $K_{ij}$ are either small or zero ($K_{ij}\approx 0$ for $i\neq j$), Eqs. (6a-b) are rewritten as,

$$-\mu\omega_t^2-\mathbf{i}\gamma\omega_t+K_{11}+\frac{G_{44}}{a^2}\big(2-2\cos(k_2 a)\big)=0, \tag{7a}$$

$$-\mu\omega_l^2-\mathbf{i}\gamma\omega_l+K_{22}+\frac{G_{11}}{a^2}\big(2-2\cos(k_2 a)\big)=0. \tag{7b}$$

For $\mathbf{P}^{\mathrm{eq}}$=($P_1^{\mathrm{eq}}$,0,0), the coefficients $K_{11}$=$\partial^2 f^{\mathrm{E}}/(\partial P_1^2)$ and $K_{22}$=$\partial^2 f^{\mathrm{E}}/(\partial P_2^2)$ are given by,

$$K_{11}=K_{11}^{\mathrm{Landau}}+K_{11}^{\mathrm{elas}}, \tag{8a}$$

$$K_{11}^{\mathrm{Landau}}=2\alpha_1+12{P_1^{\mathrm{eq}}}^2\alpha_{11}+30{P_1^{\mathrm{eq}}}^4\alpha_{111}+56{P_1^{\mathrm{eq}}}^6\alpha_{1111}, \tag{8b}$$

$$K_{11}^{\mathrm{elas}}=\frac{2Q_{11}(c_{11}-c_{12})(c_{11}+2c_{12})\left(3{P_1^{\mathrm{eq}}}^2 Q_{11}-\varepsilon_{11}\right)}{c_{11}+c_{12}}, \tag{8c}$$

$$K_{22}=K_{22}^{\mathrm{Landau}}+K_{22}^{\mathrm{elas}}, \tag{9a}$$

$$K_{22}^{\mathrm{Landau}}=2\alpha_1+2{P_1^{\mathrm{eq}}}^2\alpha_{12}+2{P_1^{\mathrm{eq}}}^4\alpha_{112}+2{P_1^{\mathrm{eq}}}^6\alpha_{1112}, \tag{9b}$$

$$K_{22}^{\mathrm{elas}}=\frac{2Q_{12}(c_{11}-c_{12})(c_{11}+2c_{12})\left({P_1^{\mathrm{eq}}}^2 Q_{11}-\varepsilon_{11}\right)}{c_{11}+c_{12}}, \tag{9c}$$

where $\varepsilon_{11}$ is the uniaxial strain along the $x_1$ direction. Rearranging Eq. (7a) yields the dispersion relation of the transverse wave ($A_1$(TO) mode), $\Delta P_1(x_2,t)$,

$$\omega_t = -\frac{\mathrm{i}\gamma}{2\mu} + \sqrt{\omega_1^2 + \frac{2G_{44}}{\mu a^2}\left(1-\cos(k_2 a)\right)}, \tag{10}$$

where the negative root is disregarded and $\omega_1 = \sqrt{K_{11}/\mu - (\gamma^2/4\mu^2)}$ is the real part frequency of the ferroelectric soft mode at $k_2 = 0$. The term $-\gamma^2/4\mu^2$ acts as a reduction in the soft mode frequency due to polarization damping $\gamma$ [38]. The imaginary part of the frequency, $\gamma/2\mu$, describes the decay rate of the mode amplitude, which also quantifies the spectral linewidth (half width at half maximum) of the mode [62]. The dispersion relation of the longitudinal wave (E(LO) mode), $\Delta P_2(x_2, t)$, can be derived by re-arranging Eq. (7b), i.e.,

$$\omega_l = -\frac{\mathrm{i}\gamma}{2\mu} + \sqrt{\omega_2^2 + \frac{2G_{11}}{\mu a^2}\left(1-\cos(k_2 a)\right)}, \tag{11}$$

where the negative root is disregarded and $\omega_2 = \sqrt{K_{22}/\mu - (\gamma^2/4\mu^2)}$ is the real part frequency of the E(LO) mode at $k_2 = 0$. Increasing $\gamma$ moderately leads to a small decrease in these frequencies. For example, the ferroelectric soft mode frequency $\omega_1$ reduces slightly from 4.14876 THz to 4.14851 THz as $\gamma$ increases from $2\times10^{-7}$ Ω·m (a value reported for $PbTiO_3$ [28]) to $8\times10^{-7}$ Ω·m. The system becomes critically damped only when $\gamma$ is orders of magnitude larger, e.g., $\omega_2 \approx 0$ at $\gamma = 1.8\times10^{-5}$ Ω·m.

Eq. (10) indicates that the dispersion relation of the ferroelectric soft mode in $BaTiO_3$ is determined by the gradient energy coefficient $G_{44}$, as also reported in [58]. $G_{44}$ is obtained by fitting the transverse optical phonon dispersion curve in the inelastic neutron experiment study of cubic $BaTiO_3$ [67] with an approximated dispersion relation at low $k$ and zero damping, given by, $\omega_t^2 = \omega_0^2 + (G_{44}/\mu)k^2$, where $G_{44} \approx 3\times10^{-11}$ $C^{-2}\cdot m^4\cdot N$. Since we are mainly focusing on the ferroelectric soft mode, $\Delta P_1(x_2, t)$, we assume an isotropic gradient coefficient in the dynamical phase-field simulations. This is achieved by rewriting Eq. (2) as (see [58]),

$$f_{\mathrm{grad}} = \frac{1}{2}G_{11}\left(P_{1,1}^2 + P_{2,2}^2 + P_{3,3}^2\right) + G_{14}\left(P_{1,1}P_{2,2} + P_{2,2}P_{3,3} + P_{1,1}P_{3,3}\right) + \frac{1}{2}G_{44}\left[P_{1,2}^2 + P_{2,1}^2 + P_{2,3}^2 + P_{3,2}^2 + P_{1,3}^2 + P_{3,1}^2\right], \tag{12}$$

with $G_{14} = G_{12} + G_{44}$, assuming that $P_{1,1}P_{2,2}=P_{1,2}P_{2,1}$, $P_{2,2}P_{3,3}=P_{2,3}P_{3,2}$, and $P_{1,1}P_{3,3}=P_{1,3}P_{3,1}$. By setting $G_{11} = G_{44} = G_0$ and $G_{14} = 0$, Eq. (10) reduces to an isotropic form of $f_{\mathrm{grad}} = \frac{1}{2}G_0(\nabla\mathbf{P})^2$. We now derive the restoring force for the polarization oscillation from the right-hand side of Eq. (1). The individual restoring force components can be calculated by linearizing Eq. (1),

$$\mu\frac{\partial^2 P_j}{\partial t^2} + \gamma\frac{\partial P_j}{\partial t} = G_0\nabla^2(\Delta P_i) - K_{ij}^{\mathrm{Landau}}\Delta P_j - K_{ij}^{\mathrm{elas}}\Delta P_j + E_i^{\mathrm{d}}, \tag{13}$$

where $G_0$ is the isotropic gradient energy coefficient, $K_{ij}^{\mathrm{Landau}} = \left.\frac{\partial^2 f_{\mathrm{Landau}}}{\partial P_i \partial P_j}\right|_{P_k^{\mathrm{eq}}, \varepsilon_{kl}^{eq}}$ and $K_{ij}^{\mathrm{elas}} = \left.\frac{\partial^2 f_{\mathrm{elas}}}{\partial P_i \partial P_j}\right|_{P_k^{\mathrm{eq}}, \varepsilon_{kl}^{\mathrm{eq}}}$ are the stiffness coefficient matrices, and $P_i^{\mathrm{eq}}$ and $\varepsilon_{ij}^{\mathrm{eq}}$ are the polarization and total strain at thermodynamic equilibrium. The right-hand side of Eq. (13) shows the four restoring force components, including the gradient term, $F_i^{\mathrm{grad}} = G_0\nabla^2(\Delta P_i)$, the Landau term $F_i^{\mathrm{Landau}} = -K_{ij}^{\mathrm{Landau}}\Delta P_j$, the elastic term $F_i^{\mathrm{elas}} = -K_{ij}^{\mathrm{elas}}\Delta P_j$, and the depolarization term $F_i^{\mathrm{d}} = E_i^{\mathrm{d}}$.

### 3.2. Analytical solution for coupled acoustic-optic phonon dispersions

The dispersion relations of the coherent polarization waves in Sec. 3.1 do not incorporate the bidirectional interaction with acoustic wave via the piezoelectric effect. This approximation is only acceptable when the frequencies and/or wavenumbers of the polarization and acoustic waves are significantly different. For completeness, here we derive the dispersion relations of the mutually coupled acoustic and polarization waves. We begin by considering a 3D bulk ferroelectric system, small perturbations of the polarization and mechanical displacement are introduced as, $\mathbf{P}(x_i,t)=\mathbf{P}^{\mathrm{eq}}+\Delta\mathbf{P}(x_i,t)=\left(P_1^{\mathrm{eq}},P_2^{\mathrm{eq}},P_3^{\mathrm{eq}}\right)+\left(\Delta P_1(x_i,t),\Delta P_2(x_i,t),\Delta P_3(x_i,t)\right)$ , $\mathbf{u}(x_i,t)=\mathbf{u}^{\mathrm{eq}}+\Delta\mathbf{u}(x_i,t)=\left(u_1^{\mathrm{eq}},u_2^{\mathrm{eq}},u_3^{\mathrm{eq}}\right)+\left(\Delta u_1(x_i,t),\Delta u_2(x_i,t),\Delta u_3(x_i,t)\right)$, respectively, where *i*=1,2,3, noting that $\varepsilon_{ij}=\frac{1}{2}\left(\frac{\partial u_i}{\partial x_j}+\frac{\partial u_j}{\partial x_i}\right)$ Assuming $E_i^{\mathrm{inc}}$=0, $E_i^{\mathrm{d}}$=0, and $E_i^{\mathrm{rad}}$=0, the polarization equation of motion (Eq. (1)) and the elastodynamic equation (Eq. (4)) can be expanded in terms of the linear perturbation terms $\Delta\mathbf{P}(x_i,t)$ **and** $\Delta\mathbf{u}(x_i,t)$,

$$\left(\mu\frac{\partial^2}{\partial t^2}+\gamma\frac{\partial}{\partial t}+\mathbf{K}+G_0\frac{\partial^2}{\partial x_j^2}\right)\Delta\mathbf{P}(x_i,t)+\mathbf{LR}\Delta\mathbf{u}(x_i,t)=0, \tag{14a}$$

$$\rho\frac{\partial^2}{\partial t^2}\Delta\mathbf{u}(x_i,t)=\left(1+\beta\frac{\partial}{\partial t}\right)\widehat{\mathbf{R}}\mathbf{C}\widetilde{\mathbf{R}}\mathbf{R}\Delta\mathbf{u}(x_i,t)-\left(1+\beta\frac{\partial}{\partial t}\right)\widehat{\mathbf{R}}\mathbf{C}\widetilde{\mathbf{R}}\mathbf{Q}\widehat{\mathbf{P}}^{\mathrm{eq}}\Delta\mathbf{P}(x_i,t), \tag{14b}$$

where the coefficient matrices $K_{ij}=\frac{\partial^2 f^{\mathrm{E}}}{\partial P_i\partial P_j}$, $L_{ijk}=\frac{\partial^2 f^{\mathrm{E}}}{\partial P_i\partial\varepsilon_{jk}}$, $\mathbf{C}$ is the bare (determined under zero-polarization) elastic stiffness tensor, $\mathbf{Q}$ is the electrostrictive tensor. The detailed expressions of the operator matrices ($\mathbf{R}$, $\widehat{\mathbf{R}}$, and $\widetilde{\mathbf{R}}$) and the $\widehat{\mathbf{P}}^{\mathrm{eq}}$ matrix are provided in Supplemental Note 1. Plugging the plane-wave solution for the polarization oscillation and the mechanical displacement oscillation (e.g., $\Delta\mathbf{u}(x_i,t)=\Delta\mathbf{u}^{\mathbf{0}}e^{\mathbf{i}(k_ix_i-\omega t)}$ and $\Delta\mathbf{P}(x_i,t)=\Delta\mathbf{P}^0e^{\mathbf{i}(k_ix_i-\omega t)}$) into Eq. (14b) gives,

$$\rho\omega^2=-(1-\mathbf{i}\beta\omega)\widehat{\mathbf{R}}(\mathbf{k})\mathbf{c}^{\mathrm{eff}}(\omega)\mathbf{R}(\mathbf{k}), \tag{15}$$

where $\mathbf{k}$=($k_1$,$k_2$, $k_3$) are wavenumbers along the three orthogonal axes. The effective elastic stiffness tensor considering the backaction from coherent polarization waves is expressed as,

$$\mathbf{c}^{\mathrm{eff}}(\omega)=\mathbf{C}\widetilde{\mathbf{R}}\left(\mathbf{Q}\widehat{\mathbf{P}}^{\mathrm{eq}}\left(-\mu\omega^2-\mathbf{i}\gamma\omega+\mathbf{K}+G_0(k_1^2+k_2^2+k_3^2)\right)^{-1}\mathbf{L}+\mathbf{I}\right), \tag{16}$$

where $\mathbf{I}$ is the identity matrix. Next, we consider a simplified 1D system where the lattice polarization $P_i$ is uniform along the $x_1$ and $x_3$ directions. Employing the discrete spatial element introduced in Sec 3.1, the plane-wave solution of the polarization oscillation and the mechanical displacement oscillation are expressed as $\Delta\mathbf{P}(x_2=na,t)=\Delta\mathbf{P}^0e^{\mathbf{i}(k_2na-\omega t)}\equiv\Delta\mathbf{P}_n$ , and $\Delta\mathbf{u}(x_2=na,t)=\Delta\mathbf{u}^0e^{\mathbf{i}(k_2na-\omega t)}\equiv\Delta\mathbf{u}_n$, where $a$=0.4 nm is the simulation cell size along the $x_2$ direction. Plugging $\Delta\mathbf{P}_n$ and $\Delta\mathbf{u}_n$ into Eqs. (14a-b) and ignoring the damping coefficients $\gamma$ and $\beta$, Eqs. (12a-b) can be rewritten in matrix form,

$$\mathbf{A}\begin{pmatrix}\Delta\mathbf{P}^0\\ \Delta\mathbf{u}^0\end{pmatrix}=\begin{pmatrix}-\mu\omega^2+\mathbf{K}+2G_0\left(1-\cos(k_2a)\right)/\mathrm{a}^2 & \mathbf{LR}(\mathbf{k})\\ \widehat{\mathbf{R}}(\mathbf{k})\mathbf{C}\widetilde{\mathbf{R}}\mathbf{Q}\widehat{\mathbf{P}}^{\mathrm{eq}} & -\rho\omega^2-\widehat{\mathbf{R}}(\mathbf{k})\mathbf{C}\widetilde{\mathbf{R}}\mathbf{R}(\mathbf{k})\end{pmatrix}\cdot\begin{pmatrix}\Delta\mathbf{P}_n\\ \Delta\mathbf{u}_n\end{pmatrix}=0. \tag{17}$$

The expressions of the $\widehat{\mathbf{R}}(\mathbf{k})$ and $\mathbf{R}(\mathbf{k})$ matrices in such 1D system are also provided in Supplemental Note 1. The mechanical boundary condition of a uniaxially strained membrane (Eqs. 3(a-c)) will influence the coefficient matrices $\mathbf{K}$ and $\mathbf{L}$. The frequency-wavenumber ($\omega$-$k_2$) dispersion relation of the total 6 branches (3 acoustic and 3 optical) of these coherent polarization and acoustic waves in a uniaxially strained single-domain $BaTiO_3$ membrane can be obtained by diagonalizing the matrix **A**. The projection weights of the dynamical components ($\Delta P_1^0$, $\Delta P_2^0$, $\Delta P_3^0$, $\Delta u_1^0$, $\Delta u_2^0$, $\Delta u_3^0$) for a particular eigenmode can be calculated by normalizing the corresponding eigenvector, after multiplying a dimensionalizing factor ($\mu$, $\mu$, $\mu$, $\rho$, $\rho$, $\rho$), where $\mu$ is the effective mass of the ferroelectric system, and $\rho$ is the mass density.

### 3.3. Dispersion relation of the band-folded transverse acoustic phonons

The domain periodicity results in a dispersion band-folding in momentum space. The two oppositely poled ferroelectric domains, despite having piezoelectric tensors of different signs, have an identical speed of sound $v$. Since the mass density $\rho$ is the same in these two domains, there is no mismatch in the acoustic impedance ($Z=\rho v$). As a result, the crossing at the mini-zone boundary should be gapless [68], if omitting the influence of the thin Ising wall itself with spatially varying polarization. Quantitatively, let us consider the propagation of a transverse acoustic (TA) wave in a 1D system with equilibrium polarization along the $\pm x_1$ axis. The domain periodicity (i.e., the width of a single domain or the distance between the centers of two neighboring domains along the $x_2$ axis [69]) is denoted as $D$. The velocity of TA wave is $v_t = \sqrt{c_{44}^{\mathrm{eff}}/\rho}$, where $c_{44}^{\mathrm{eff}}$ is the frequency-dependent elastic stiffness coefficient (see Eq. (16)) that is identical in the two oppositely poled domains. Using the discretized coordinate $x_2 = na$, the plane-wave solution of the two counter-propagating TA waves can be expressed as,

$$u_{1,n}(t) = u_{1,n}^{+} e^{\mathbf{i}(k_t na - \omega_t t)} + u_{1,n}^{-} e^{\mathbf{i}(-k_t na - \omega_t t)}, \tag{18}$$

where $k_t$ is the angular wavenumber of the TA wave in each domain period and $\omega_t$ is the corresponding angular frequency. In a 1D system that only hosts a TA wave, $\Delta u_1(x_2, t)$, the only nonzero stress component is $\sigma_{12}$, given by,

$$\sigma_{12,n} = \frac{c_{44}^{\mathrm{eff}}(u_{1,n+1} - u_{1,n})}{a}. \tag{19}$$

Ignoring the elastic damping ($\beta$=0), the elastodynamic equation (Eq. (4)) reduces to,

$$-\rho\omega_t^2 = \frac{c_{44}^{\mathrm{eff}}}{a^2}\left(e^{ik_t a} + e^{-ik_t a} - 2\right). \tag{20}$$

Rearranging Eq. (19) yields the TA dispersion within a uniform domain,

$$\omega_t = \sqrt{\frac{2c_{44}^{\mathrm{eff}}}{\rho a^2}\left(1 - \cos(k_t a)\right)}. \tag{21}$$

The periodic boundary conditions of both displacement $u_1$ and stress $\sigma_{12}$ are expressed as [70]:

$$u_{1,n}(na = 0) = u_{1,n}(na = d)e^{\mathbf{i}K_t D}, \tag{22a}$$

$$\sigma_{12,n}(na = 0) = \sigma_{12,n}(na = d)e^{\mathbf{i}K_t D}, \tag{22b}$$

where $K_t$ is the effective (superlattice) wavevector which defines the folded Brillouin zone $-\pi/D \le K_t \le \pi/D$. Substituting Eqs. (18,19) into Eq. (20) yields the equations in matrix form,

$$\begin{bmatrix} 1 - e^{\mathbf{i}D(K_t + k_t)} & 1 - e^{\mathbf{i}D(K_t - k_t)} \\ \frac{c_{44}^{\mathrm{eff}}}{a}\left(e^{iak_t} - 1\right)\left(1 - e^{iD(K_t + k_t)}\right) & \frac{c_{44}^{\mathrm{eff}}}{a} e^{-i(a+D)k_t}\left(e^{iak_t} - 1\right)\left(e^{iDK_t} - e^{iDk_t}\right) \end{bmatrix} \begin{bmatrix} u_1^{+} \\ u_1^{-} \end{bmatrix} = 0. \tag{23}$$

Obtaining non-trivial solution for Eq. (23) requires that the determinant of the coefficient matrix to be zero, which gives rise to the resonant condition for the TA waves within the first lattice Brillouin zone $-\pi/a \le k_t \le \pi/a$,

$$\sin(\frac{D}{2}(K_t - k_t))\sin(\frac{D}{2}(K_t + k_t)) = 0, \quad k_t = \pm K_t + n\frac{2\pi}{D}, n \in \mathbb{Z} \tag{24}$$

Thus, the dispersion relation of the band-folded TA waves is given by,

$$\omega_t = \sqrt{\frac{2c_{44}^{\mathrm{eff}}}{\rho a^2}\left(1 - \cos\left(K_t + n\frac{2\pi}{D}\right)a\right)}, \tag{25}$$

where integer $n$ indicates multiple band-folding in the momentum space. We note that the dispersion relations above show that all frequencies are allowed, suggesting the absence of frequency band gaps. This is consistent with the fact that the there is no acoustic impedance mismatch between two adjacent domains that have the same sound velocity and mass density.

## 4. Intrinsic modes of coherent polarization waves

We first study the intrinsic modes of coherent polarization waves. A single-domain, uniaxially strained $BaTiO_3$ membrane with a spontaneous polarization along $+x_1$ is shown in Fig. 1(a). The single-domain state is obtained by performing 3D dynamical phase-field simulations under a constant temperature of 300 K and constant total strain $\varepsilon_{11}$ of 0.7%, where the local polarization vectors in the initial state ($t$=0) all align along $+x_1$ but have a small magnitude fluctuating near 0.01 C/m$^2$. This initial state mimics the paraelectric state. The membrane undergoes an isothermal paraelectric-to-ferroelectric phase transition, forming a single domain at equilibrium. Specifically, all the local polarization evolves from a near-zero but positive value to its nearest minimum at $P_1^{\mathrm{eq}}$=0.26 C/m$^2$ in the profile of the local free energy density $f^{\mathrm{E}}$, as shown in Fig. 1(b). Figure 1(c) shows the polarization oscillation in the center of the membrane, $\Delta P_1(t)=P_1-P_1(t=0)$. The corresponding frequency spectrum is shown in Fig. 1(d), which shows a frequency band near 4-6 THz with the lowest frequency of approximately 4.16 THz.

In the initial stage of domain formation, the initial small-amplitude polarization vectors rapidly grow towards $P_1^{\mathrm{eq}}$, manifesting as the growth of polarization wave [71], which is analogous to the growth of concentration wave in spinodal decomposition. Figure 1(e) shows the profiles of transverse polarization wave $\Delta P_1(x_2,t)$ along the dashed line indicated in Fig. 1(a) at several representative moments of the initial growth stage. Performing 2D Fourier transform of the $\Delta P_1(x_2,t)$ gives the dispersion relation of the transverse polarization wave represented by $\Delta\tilde{P}_1(k_2, \omega)$ in Fig. 1(f), where $k_2$ and $\omega$ are the angular wavenumber and angular frequency, respectively. The temporal axis of the Fourier transform considered the period from $t$=30 to 200 ps. The polarization evolution in this period exhibits the oscillations around $P_1^{\mathrm{eq}}$ that are apparent in Fig. 1(c) and is thus the most insightful range for the dispersion relation analysis.

Figure 1(f) conveys three main messages. First, the frequency band within 4-6 THz shows the largest spectral amplitude, which is consistent with the observation in Fig. 1(d). The frequency band shows a parabolic dependence on the wavevector near $k_2$=0, and a saturation near $k_2/2\pi=1/(2a)$ due to the discrete spatial grids. Here, $a$=0.4 nm is the size of simulation cell along the $x_2$ axis, which sets the upper-bound of the wavenumber. The $\omega$-$k_2$ relation within this frequency band agrees well with the analytically calculated dispersion relation for the transverse polarization wave, $\Delta P_1(x_2,t)$, in uniaxially strained $BaTiO_3$ membrane (see Eq. (10)), which corresponds to the ferroelectric soft mode (the transverse optical $A_1$(TO) mode [72]), as shown by the dashed line in Fig. 1(f). At $k_2$=0, the frequency of the $A_1$(TO) mode is $\omega_1/2\pi$=4.15 THz. This value depends on the curvature of the free energy landscape, quantified by $\partial^2 f^{\mathrm{E}}/\partial P_1^2$ at $P_1=P_1^{\mathrm{eq}}$.

Second, there also exists an appreciable frequency band of similar feature within 1-4 THz, which agrees well with the analytically calculated dispersion relation for the longitudinal polarization wave $\Delta P_2(x_2,t)$ (see Eq. (11) in Sec. 3.1), which corresponds to a longitudinal optical E(LO) mode [66] with a temporal frequency of $\omega_2/2\pi$=1.10 THz at $k_2$=0. $\omega_2$ differs from $\omega_1$ because of the different local curvature, quantified by $\partial^2 f^{\mathrm{E}}/\partial P_2^2$ at $P_1$=$P_1^{\mathrm{eq}}$. Notably, the fact that the spectral feature of $\Delta P_2(x_2,t)$ can be seen from the numerically simulated $\Delta\tilde{P}_1(k_2, \omega)$ suggests the anharmonic coupling between the TO and LO phonons, which mainly results from the high-order term in the effective electric field from the Landau free energy density [38].

Third, there exist other modes with much weaker spectral amplitudes. These polarization wave modes may be induced by acoustic waves via the piezoelectric coupling and are therefore called *acoustic* modes. Nonlinear mixing among different coherent polarization waves, e.g., sum/difference frequency generation, may also contribute to these weaker modes.

To provide a clean understanding of all the possible modes in the present membrane system and exclude the nonlinear modes, we analytically calculate the dispersion relations for the coupled acoustic waves (acoustic phonons) and polarization waves (optical phonons) by solving the linearized equations of motion for the polarization and mechanical displacement in the frequency domain. Details of the analytical calculation are shown in Sec. 3.2. The calculation results are shown in Fig. 1(g), which contains two optical branches corresponding to $\Delta P_1$ ($A_1$(TO) mode) and $\Delta P_2$ (*E*(LO) mode) as well as all three acoustic branches corresponding to $\Delta u_1$, $\Delta u_2$, and $\Delta u_3$. The $\Delta P_3$ branch has frequencies above 20 THz due to the large out-of-plane depolarization field $E_3^{\mathrm{d}}$ [41] and thus is not shown in the figure. Notably, the $\Delta P_2$ branch (orange) is superimposed with a small portion of the $\Delta u_1$ projection (blue), and vice versa, indicating a bidirectional piezoelectric coupling between the longitudinal polarization wave component $\Delta P_2(x_2,t)$ and the transverse acoustic wave component $\Delta u_1(x_2,t)$ in the present system. Such $\Delta P_2$-$\Delta u_1$ coupling is also evident from the bar plots in Fig. 1(h), which show the normalized weighted projections of the $\Delta P_i$ and $\Delta u_i$ components for the six frequency modes at $k_2/2\pi$=0.5 nm$^{-1}$.

## 5. Coherent polarization waves confined to domain walls

We now turn to studying the DW modes by extracting the dispersion relation of coherent polarization waves in stripe domains. The striped domains containing Ising-type 180° DWs is obtained by setting $P_i(\mathbf{r}, t=0)$ to be along the $+x_1$ or $-x_1$ axis, with $P_1\approx\pm0.01$ C/m$^2$ and $P_2$=$P_3$=0, in eight prescribed regions and letting the system evolve to equilibrium. The simulations employed a constant temperature of 300 K and a constant $\varepsilon_{11}$ of 0.7%. The polarization charges are screened by mobile charges at all four free surfaces ($x_1$-$x_3$ and $x_1$-$x_2$ surfaces). Figure 2(a) shows the equilibrium ($t$=0) multidomain structure and the corresponding line profile of $P_1$ is shown in Fig. 2(b). The polarization evolution at point I (in the domain center) is shown in Fig. 2(c). Figure 2(d) shows the polarization evolution at point II/III, which are near/at the DW center, respectively.

Figures 2(e,f) show the frequency spectra obtained using Fourier transform of the time-domain data in Figs. 2(c,d), respectively, both for the period of $t$=30-200 ps. The spectral feature of the point at the domain center (Fig. 2(e)) is similar to the single-domain case in Fig. 1(c). As shown in Fig. 2(e), the most significant mode at the domain center is the ferroelectric soft mode, or $A_1$(TO) mode, at the frequency band from 4-6 THz. The minor frequency peak around 1.36 THz should be associated with the E(LO) mode. Figure 2(f) contains three major peaks at around 0.22 THz, 3.01 THz, and 3.49 THz, which are all below the soft-mode frequencies. These lower-frequency modes are likely the DW modes because the effective mass of a DW is expected to be larger than the effective mass of an optical phonon. In particular, the 3.01 THz peak, which appears in the spectrum of point II, vanishes in the spectrum of point III at the DW center. As will be elaborated later, this feature indicates a DW breathing mode [53,73] where the location of DW center is fixed.

Figure 2(g) further shows the 2D Fourier transform of the polarization oscillation $\Delta P_1(x_2, t)$ along the dashed line indicated in Fig. 2(a). Figure 2(g) conveys three messages. First, the frequency bands at 4-6 THz and 1.5-4 THz, which are parabolic at low $k_2$ and asymptotic at high $k_2$, correspond to the dispersion curves of $A_1$(TO) and E(LO) modes, respectively. Second, there exist three distinct, non-dispersive bands below the soft mode frequencies, which are consistent with the emergence of three peaks at 0.22 THz, 3.01 THz, and 3.49 THz in Fig. 2(f). These non-dispersive bands are therefore likely to arise from the DW modes. Third, dense sets of modes also appear over a wide range of frequency and wavenumber. These modes correspond to the polarization waves induced via the piezoelectric coupling by the band-folded cavity TA phonons, $\Delta u_1(x_2,t)$, which form due to the simultaneous presence of periodic domains and the two free surfaces along the $x_2$ axis.

These dense sets of modes are more evident in the $\Delta\tilde{P}_2(k_2, \omega)$ spectrum (Fig. 2(h)), which is likely because the $\Delta P_2$-$\Delta u_1$ coupling is the only allowable type of coupling between the acoustic and polarization waves in the present $BaTiO_3$ membrane that is periodically poled along $x_1$ (see Fig. 1(h)). For further clarification, Figure 2(i) shows the analytically calculated dispersion relations of the band-folded TA phonons $\Delta u_1(x_2,t)$; see details in Sec. 3.3. The dispersion curves for the $A_1$(TO) and E(LO) modes are also plotted. The wavenumbers of these band-folded acoustic phonons are determined by the domain periodicity $D\approx 16$ nm, with $k_2/2\pi=n/2D$. The upper frequency bound for the TA phonons $\omega_t^{\max} = \sqrt{\frac{c_{44}^{\mathrm{eff}}}{\rho}}\frac{\pi}{a}$, where $c_{44}^{\mathrm{eff}}$ is the frequency-dependent elastic stiffness coefficient (see Eq. (16) in Sec. 3.2). As shown in Fig. 2(i), the dispersion curve of the E(LO) mode polarization wave, $\Delta P_2(x_2, t)$, overlaps with the dispersion curves of the band-folded TA phonons over a wide range of frequencies and wavenumbers. As a result, the E(LO) mode would be resonantly excited. This is analogous to the resonant interaction between acoustic phonons and magnons via the magnon-phonon coupling [74-76]. Furthermore, the two free surfaces along the $x_2$ axis enable the formation of a cavity for both $\Delta P_2(x_2, t)$ and $\Delta u_1(x_2,t)$, which extends their interaction time and leads to cavity-enhanced resonant excitation of polarization wave. By contrast, the $A_1$(TO) mode cannot be resonantly coupled to the TA phonons due to the absence of $\Delta P_1$-$\Delta u_1$ coupling in the present system. For the same reason, the DW modes, which predominantly contain the $\Delta P_1(x_2,t)$ component, could not interact resonantly with TA phonons, although their frequencies can match. This is evidenced by the absence of the DW mode features in the $\Delta\tilde{P}_2(k_2, \omega)$ spectrum in Fig. 2(h).

## 6. Near-equilibrium excitation and profile reconstruction of domain wall modes

The dispersion relations discussed above are obtained from Fourier transform of the spatiotemporal evolution of polarization in the last stage of ferroelectric domain formation. Such simulation is straightforward to implement and allows for accessing a broad range of polarization wave modes, but it is challenging to exclude modes resulting from the nonlinear interaction among different modes, making it difficult to distinguish the DW modes from the other modes even in the present system with simple 180º Ising-type DWs.

In this regard, we further apply a broadband THz pulse to the equilibrated stripe domain shown in Fig. 2(a). The THz pulse has a relatively small peak amplitude (2 kV/cm, see Fig. 3(a)) to ensure a weak perturbation of the system and has a center frequency of 1.5 THz with a full-width-half-maximum bandwidth of 2.55 THz (Fig. 3(b)) to cover most of the polarization wave modes discovered in Fig. 2(g). Figure 3(c) shows the polarization oscillation $\Delta P_1$ for $t$=50-200 ps at two points in the DW region, as indicated by red and blue colors in the inset, corresponding to points II and III in Fig. 2(b). Within 50-200 ps, the applied electric field has become almost zero, thus the polarization oscillation should not result from field-induced oscillation. The frequency spectra in Fig. 3(d), which are obtained from Fourier transform of the time-domain data in Fig. 3(c), contains four strong peaks at 0.17 THz, 2.94 THz, 3.35 THz and 4.17 THz. Based on the analytically calculated soft-mode dispersion relation (Eq. (10)), the 4.17 THz peak characterizes the $k_2$=0 $A_1$(TO) mode. The 2.94 THz peak is not visible in the spectrum of $\Delta P_1$ oscillation at the DW center (point III), consistent with the nature of a DW breathing mode.

Figure 3(e) shows the spatially resolved frequency spectrum $\Delta\tilde{P}_1(x_2, \omega)$, obtained by time-domain Fourier transform for the $\Delta P_1(x_2,t)$ along the same dashed line indicated in Fig. 2(a), for the period of $t$=50-200 ps followed by the broadband THz pulse excitation. As shown in Fig. 3(e), the peaks near 0.17 THz, 2.94 THz, and 3.35 THz are all localized at the DW regions, indicating their nature as DW bounded states with negligible interaction among different DWs and that all DWs are collectively excited. Figure 3(f) further shows the $\Delta\tilde{P}_1(k_2, \omega)$ spectrum obtained by 2D Fourier transform of $\Delta P_1(x_2,t)$. Comparing Fig. 3(f) to Fig. 2(g), one can see that the polarization wave modes induced by the band-folded acoustic phonons are still evident but mostly below 1 THz. Moreover, the DW modes near 0.17 THz, 2.94 THz, and 3.35 THz are more distinct. As $\gamma$ increases from $2\times10^{-7}$ to $8\times10^{-7}$ Ω·m (see control simulations in Supplemental Fig. S1), the 0.17 THz DW mode shows a slight decrease in resonant frequency (from 0.17 THz to 0.156 THz) along with a noticeable increase in the linewidth. The effects of $\gamma$ on both the frequency and linewidth of the 2.94 THz, and 3.35 THz DW modes are not apparent likely because the damping-induced change is too small compared to their natural frequencies. This is similar to the small effect of $\gamma$ on the frequency and linewidth of ferroelectric soft mode within this range of $\gamma$, as discussed in the text below Eqs. (10,11).

To quantitatively understand the DW modes of an Ising-type 180º DW, we extend the 1D numerical eigenmode analysis that was reported for ferroelectric bulk crystals [53] and strained thin films [73] to the present case of a uniaxially strained membrane. The Ising-type 180º DW remains nearly translational invariant within the wall plane (e.g., along the $x_1$ and $x_3$ axes), and the dominant dynamical degree of freedom is associated with the polarization variations along the DW normal (e.g., the $x_2$ axis). Thus, the essential DW dynamics can be adequately captured by the 1D numerical eigenmode analysis. Details of our eigenmode analysis are shown in Appendix A. The first-order eigenmode of the DW is a zero-energy (Goldstone) sliding mode. The second-order eigenmode is a DW breathing mode that has a much higher frequency of 2.69 THz. The third-order eigenmode has a frequency of 3.20 THz. The mode profiles predicted from the eigenmode analysis are also shown in Appendix A.

With these in mind, we first reconstruct the spatiotemporal evolution of the DW polarization profile via the inverse Fourier transform for the 2.94 THz peak, as shown in Fig. 3(g). Detailed data processing procedures are provided in Appendix B. The reconstructed profiles indicate a well-defined DW breathing mode, where the single grid at the DW center remains unperturbed, local polarization at the two neighboring grids next to the DW center shifts towards (away from) the horizontal line of $P_1$=0 during the expansion (reduction) of the DW thickness. Next, we reconstruct the 1D DW profiles for the 3.35 THz peak. In contrast with the breathing mode at 2.94 THz, its spatial profile becomes asymmetric with respect to the DW center at equilibrium, as shown in Fig. 3(h). Such an asymmetric profile resembles the third-order eigenmode at 3.20 THz (see Appendix A). The DW profiles reconstructed for the 0.17 THz peak, as shown in Fig. 3(i), appear to be the same as the zero-energy sliding mode (see Appendix A). This raises the question of the origin of the restoring force for this unconventional DW sliding mode. In particular, unlike the case in [77] where the surface-polarization-charge induced depolarization field provides a restoring force, there are no bound polarization charges at the two $x_2$-$x_3$ surfaces herein due to the use of periodic boundary condition along the $x_1$ axis.

To understand the origin of such unconventional DW sliding mode, we reconstruct the 2D spatiotemporal distributions of polarization oscillation $\Delta\mathbf{P}$ at 0.17 THz via the inverse Fourier transform, as shown in Fig. 4(a). The corresponding distributions of bulk polarization bound charge, denoted as $\rho_{\mathrm{b}} = -\nabla\cdot(\Delta\mathbf{P})$, show appreciable non-uniformity along the $x_1$ axis, as shown in Fig. 4(b). Such nonzero local bound charges are primarily contributed by the spatially varying $\Delta P_1$ ($x_1$,$t$), i.e., $-\nabla\cdot(\Delta\mathbf{P}) \approx -\partial\Delta P_1/\partial x_1$ (see Supplemental Fig. S2), which arises from the intrinsic LO phonon mode. These bulk non-uniform bound polarization charges produce a depolarization field along the $x_1$ axis, shown in Fig. 4(c), which acts as a restoring force for DW sliding along the $x_2$ axis. Figure 4(d) shows the local DW displacement with respect to the initial equilibrium location, $\Delta x_2$, along the $x_1$ axis at different time stages corresponding to those in Fig. 4(a). As shown, the magnitude of $\Delta x_2$ varies slightly along the $x_1$ axis at each time stage, demonstrating a wavy profile of the DW during sliding. Moreover, at 50 ps and 51 ps, a positive $\Delta P_1$ leads to a negative $\Delta x_2$, and vice versa for the case at 52 ps and 53 ps. This trend is consistent with the 1D line profiles shown in Fig. 3(i). Figure 4(e) further shows the evolution of the reconstructed $\Delta P_1$(t) at the DW center ($x_1$, $x_2$)=(90 nm, 34 nm). The corresponding evolution of the depolarization field component, $E_1^d$, and the sum of the remaining restoring forces (Landau, elastic, gradient, see their expressions in Eq. (13)), $F_1^{\mathrm{other}}$, is shown in Fig. 4(f). As seen, the peak amplitudes of $E_1^{\mathrm{d}}$ and $F_1^{\mathrm{other}}$ are comparable with $E_1^{\mathrm{d}}$ being slightly larger, suggesting that $E_1^{\mathrm{d}}$ plays an essential role in the restoring force of the unconventional DW sliding mode.

**7. Strain modulation of domain wall modes**

We now turn to discussing the effect of strain on both the intrinsic polarization wave and the DW modes. We perform dynamical phase-field simulations using the same set-up as those in Fig. 3 but vary the uniaxial strain $\varepsilon_{11}$ from 0.5% to 1.1%. Based on the temperature-strain stability diagram we computed for $BaTiO_3$ membrane (see Supplemental Fig. S3 and Note 2), $\varepsilon_{11}$ of 0.5% and above is required to stabilize a tetragonal phase $BaTiO_3$ with 180º DWs at 300 K. Moreover, DW becomes narrower as $\varepsilon_{11}$ increases, making it more difficult to resolve using a 0.4-nm simulation cell. In view of this, $\varepsilon_{11}$ is kept below 1.1%. Figure 5(a) shows the 1D profiles of the 180º DWs simulated under different strains, where the magnitude of $P_1^{\mathrm{eq}}$ in the interior of a domain increases as $\varepsilon_{11}$ increases. At $\varepsilon_{11}$=0.5% and 0.7%, there exists one data point at the DW center. This is critical to representing the DW breathing mode, which has a fixed DW center position (Fig. 3(g)). At $\varepsilon_{11}$=0.9% and 1.1%, there is no longer a data point at the DW center, making it challenging to accurately calculate the frequency/energy of the DW breathing mode. Similarly, resolving the higher-order DW eigenmode (Fig. 3(h)) would require an even smaller simulation cell. However, because the currently used simulation cell size (0.4 nm) is already near the unit cell size of $BaTiO_3$, using smaller cell sizes would lead to an unphysically large gradient energy and contradict the continuum assumption of a phase-field model.

Figure 5(b) shows the effect of $\varepsilon_{11}$ on the frequencies of the intrinsic $A_1$(TO) mode at $k_2$=0 as well as the three DW modes discussed in Figs. 3(g-i). The frequencies of the intrinsic $A_1$(TO) mode agree well with analytical calculation. The increased frequency under larger $\varepsilon_{11}$ results from the larger curvature of the free energy landscape in the domain. The frequencies of the DW breathing mode increase with $\varepsilon_{11}$ in the range of 0.5% to 0.9%. This can be attributed mainly to strain-enhanced local curvature of the free energy landscape across the DW, as indicated by Eq. (A1) in Appendix A. However, the frequency drops when $\varepsilon_{11}$ increases to 1.0% and then increases again as $\varepsilon_{11}$ further increases to 1.1%. This anomalous feature arises because the lack of a data point at the DW center (see Fig. 5(a)) makes it challenging to accurately simulate the DW breathing mode. Similarly, while we see strain-induced almost linear increase in the higher-order DW eigenmode in the range of $\varepsilon_{11}$=0.5-0.7%, abrupt increase in frequency occurs when $\varepsilon_{11}$ increases further to 0.8%, which is most likely to occur because of the fundamental cell size limit of the mesoscale phase-field model. The fact that this anomaly emerges at a lower strain (0.8%) than the case of DW breathing mode (1.0%) is consistent with the expectation that finer simulation cells are required to resolve the higher-order DW eigenmode.

The frequencies of the unconventional DW sliding mode, $\omega_{\mathrm{DW}}$, remain largely unchanged in the range of $\varepsilon_{11}$=0.5%-0.8% before starting to increase at larger values of $\varepsilon_{11}$ (0.9%-1.1%). We attribute the relatively weak strain sensitivity in the range of 0.5%-0.8% to the competing effects on the strain-modulation of $E_1^{\mathrm{d}}$ and other restoring forces. Specifically, our analyses for $\varepsilon_{11}$=0.7% (Fig. 4) have shown that the depolarization field $E_1^{\mathrm{d}}$ and other restoring forces are comparable in amplitudes with complex interplay. To understand the evident increase of $\omega_{\mathrm{DW}}$ at larger strain values, we likewise plot the evolution of $E_1^{\mathrm{d}}$ and the sum of the remaining restoring forces, $F_1^{\mathrm{other}}$, based on the $\Delta P_i(t)$ at the DW center ($x_1$, $x_2$)=(90 nm, 34 nm) reconstructed at the specific values of $\omega_{\mathrm{DW}}$. Figures 5(c,d) show the evolution of $E_1^{\mathrm{d}}$ and $F_1^{\mathrm{other}}$ for $\varepsilon_{11}$=0.9% and $\varepsilon_{11}$=1.1%, with $\omega_{\mathrm{DW}}/2\pi$=0.43 THz and 0.90 THz, respectively. In contrast with the case of $\varepsilon_{11}$=0.7% (Fig. 4(e)), the peak amplitude of $E_1^{\mathrm{d}}$ is about two orders of magnitude larger than the peak amplitude of $F_1^{\mathrm{other}}$ in both cases, indicating a dominant contribution from $E_1^{\mathrm{d}}$.

Based on this finding, we develop a simplified 1D analytical model (see Appendix C) to quantitatively understand the effect of the depolarization field on $\omega_{\mathrm{DW}}$ at large strains. Our analytical model predicts that $E_1^{\mathrm{d}}$ is inversely proportional to the DW width (denoted as $W$), which is qualitatively consistent with the simulation results in Fig. 5(c-d). Physically, a reduction in the DW width (which occurs under large $\varepsilon_{11}$) concentrates the bulk bound charge into a narrower region, thereby increasing the charge density and enhancing the depolarization field [78]. This analytical model also allows for estimating the effective mass density of the DW, $\mu_{\mathrm{DW}}$, based on the numerically extracted $\omega_{\mathrm{DW}}$, as well as the simulated equilibrium DW profile. Specifically, at $\varepsilon_{11}$=0.9%, using $P_1^{\mathrm{eq}}$=0.27 C/m$^2$, $W \approx 0.81$ nm, and $\omega_{\mathrm{DW}}/2\pi$=0.43 THz, one has $\mu_{\mathrm{DW}} = 5.26\times10^{-7}$ kg/m$^2$. At $\varepsilon_{11}$=1.1%, with $P_1^{\mathrm{eq}}$=0.30 C/m$^2$, $W \approx 0.75$ nm, and $\omega_{\mathrm{DW}}/2\pi$=0.9 THz, the same procedure yields $\mu_{\mathrm{DW}} = 1.36\times10^{-7}$ kg/m$^2$. The lower $\mu_{\mathrm{DW}}$ at larger strain is consistent with the reduced wall thickness $W$.

## 8. Conclusions

In summary, we performed dynamical phase-field simulations of the coupled polarization-strain dynamics during the formation of striped ferroelectric domains with Ising-type 180° domain walls in uniaxially strained ferroelectric membranes, using $BaTiO_3$ as an example. From the simulation results, we identified both the intrinsic polarization waves, which correspond to optical phonon modes, $A_1$(TO) and E(LO), as well as polarization waves confined to domain walls (DW modes). Based on subsequent dynamical phase-field simulations of near-equilibrium excitation by a single-cycle broadband THz pulse, we selectively excite and numerically reconstructed the spatiotemporal profiles of the individual DW modes.

Specifically, we unravel and reconstruct the profiles of canonical eigenmodes of a 180° DW, including the breathing mode and the higher-order mode, both having a frequency in the THz regime, which is consistent with previous reports [53,73]. Notably, we discover an unconventional DW sliding mode which has a sub-THz resonant frequency that arises mainly due to the uncompensated bulk polarization charges (Fig. 4). This unconventional DW sliding mode therefore differs from both the zero-energy (Goldstone) sliding mode (i.e., the lowest-order eigenmode [53]) and surface-polarization-charge induced sliding mode [50,77] of 180° DWs. Using a combination of analytical calculation, 1D eigenmode analysis, and dynamical phase-field simulations, we demonstrated that the frequencies of both the intrinsic polarization waves and these DW modes can be effectively tuned by strain via the modulation of the underlying storing forces (see Sec. 3.1).

Experimentally, we expect that these modes can be excited by a single-cycle broadband THz pulse (e.g., the condition simulated in Fig. 3) and probed by X-ray or optical pulses, similarly to the experimental discovery of the collective modes of topological polar textures in strained $PbTiO_3/SrTiO_3$ thin-film superlattices [28,31,37,79]. Furthermore, although the periodic domain patterns with eight domains are considered as an example in the simulations (Figs. 2 and 3), the excitation of the predicted DW modes does not necessarily require a perfectly periodic domain pattern or a specific number of domain periods (see control simulation in Supplemental Fig. S4) due to highly localized nature of these modes (Fig. 3(e)). Rather, any ferroelectric membranes having 180° stripe domains, such as the uniaxially strained $SrTiO_3$ membrane [19], can be used to validate the predictions in this work.

Overall, our findings provide new fundamental insights into the nature of the collective excitation of ferroelectric DWs. From an application perspective, the ability to tune DW mode frequencies by an applied strain establishes a pathway towards an on-demand control of resonant phonon-DW coupling in the THz range, which can potentially be utilized to design reconfigurable THz and optical devices. Moreover, our theoretical analyses and numerical approach can be readily extended to discover new collective modes of other types of DWs in ferroelectric materials, such as striped domains with 90° or 109° DWs (see Supplemental Fig. S3), or 180° DWs with mixed Bloch-Néel-Ising Character [80]. The fundamental understanding established through these relatively simple domain patterns will benefit the search for more complex yet intriguing collective modes from a wide range of topologically nontrivial polar textures [81-84]. Furthermore, by analogy to the strong magnon-phonon coupling in magnetic multilayers [85-88] and freestanding magnetic membranes [76], it would be possible to create new hybridized states via cavity-enhanced strong coupling between coherent polarization waves and acoustic waves [62]. These new hybridized states can potentially be utilized to design new hybrid dynamical and quantum systems [89] for realizing new functionalities in quantum communication, computing, and sensing.

**Acknowledgements**

The dynamical phase-field simulations in this work are primarily funded by the National Science Foundation (NSF) under Grant No. DMR-2237884 (X.G. and J.-M.H.) and partially funded by the Wisconsin MRSEC (DMR-2309000) (X.G., J.W., and J.-M.H.). The dynamical phase-field simulations were performed using Bridges at the Pittsburgh Supercomputing Center through allocation TG-DMR180076 from the Advanced Cyberinfrastructure Coordination Ecosystem: Services & Support (ACCESS) program, which is supported by NSF Grants No. 2138259, No. 2138286, No. 2138307, No. 2137603, and No. 2138296. The theoretical works on the analytical calculation of domain wall modes are funded by the US Department of Energy, Office of Science, Basic Energy Sciences, under Award Number DE-SC0020145 as part of the Computational Materials Sciences Program (Y.Z., L.-Q.C., and J.-M.H.). PGE acknowledges support from the U.S. Department of Energy, Basic Energy Sciences, under Contract No. DE-FG02-04ER46147. A.R. acknowledges the support of the National Science Foundation Graduate Research Fellowship Program under Grant No. DGE1255832.

**Appendix A: 1D numerical analysis of the eigenmodes of the 180° Ising domain walls**

We consider a 1D Ising-type 180° DW under the conditions of $\lim_{x_2\to\pm\infty} P_1(x_2) = \pm|P_1^{0,\mathrm{eq}}|$, where $|P_1^{\mathrm{eq}}|$ is the magnitude of equilibrium polarization in the domain center and can be determined by minimizing the Helmholtz free energy density $f^{\mathrm{E}}$. The equilibrium polarization profile across the DW, $P_1^{\mathrm{eq}}(x_2)$, can be obtained by solving the 1D Euler-Lagrange equation, $\delta F_{\mathrm{tot}}/\delta P_1 = 0$.

The numerically solved profile, $P_1^{\mathrm{eq}}(x_2)$, is shown in Fig. 6(a-c), and can be fitted using $P_1^{\mathrm{eq}}(x_2) = |P_1^{0,\mathrm{eq}}|\varphi\left(\frac{x_2-X}{W}\right)$, where $X$ denotes the coordinates of the DW center and $W$ is the DW width. The shape function $\varphi(\xi)$ satisfies the conditions of $\lim_{\xi\to\pm\infty}\varphi(\xi) = \pm 1$ and $\varphi(0) = 0$. Dynamical evolution of the DW profile, $P_1(x_2, t)$, is governed by the equation of motion for the polarization [Eq. (1)]. Substituting $P_1(x_2,t)$=$P_1^{\mathrm{eq}}(x_2) + \Delta P_1(x_2,t)$ into Eq. (1), setting $\gamma$=0, and omitting the depolarization field, Eq. (1) can be rewritten into a linearized perturbation form,

$$\mu\frac{\partial^2 \Delta P_1(x_2,t)}{\partial t^2} = \left[G_0\frac{\partial^2}{\partial x_2^2} - K_{11}^{\mathrm{Landau}}(x_2) - K_{11}^{\mathrm{elas}}(x_2)\right]\Delta P_1(x_2,t) \equiv F_1^{\mathrm{other}}, \tag{A1}$$

which resembles Eq. (13) in Sec. 3.1, except that the stiffness coefficients $K_{11}^{\mathrm{Landau}}$ and $K_{11}^{\mathrm{elas}}$ herein are functions of the equilibrium polarization values across the DW, $P_1^{\mathrm{eq}}(x_2)$, which are spatially varying. Increasing the strain $\varepsilon_{11}$ will modulate $P_1^{\mathrm{eq}}(x_2)$, leading to a larger $|P_1^{0,\mathrm{eq}}|$ in the domain center and a narrower DW width (hence the second-order spatial derivatives). Thus, strain can modulate the restoring force $F_1^{\mathrm{other}}$ by tuning all its three constituent restoring forces.

The solution to Eq. (A1) can be expanded in terms of the normal modes $\psi_n(x_2)$,

$$\Delta P_1(x_2,t) = \sum_n a_n\psi_n(x_2)e^{-\mathrm{i}\omega_n t}, \tag{A2}$$

where $\psi_n(x_2)$ and resonant frequencies $\omega_n$ are, respectively, the eigenfunctions and the eigenfrequencies that can be obtained from solving the eigenvalue problem, i.e.,

$$\left[G_0\frac{\partial^2}{\partial x_2^2} - K_{11}^{\mathrm{Landau}}(x_2) - K_{11}^{\mathrm{elas}}(x_2)\right]\psi_n(x_2) = -\omega_n^2\psi_n(x_2), \tag{A3}$$

To numerically solve Eq. (A3), the operator in the bracket is discretized along the $x_2$ axis using a uniform cell size of $\Delta x_2$=0.4 nm, the same as the dynamical phase-field simulation. The second-order spatial derivative is approximated by a centered finite-difference scheme, resulting in a tridiagonal real-symmetric matrix $\boldsymbol{\mathcal{H}}$. Eq. (A3) is then transformed into a standard matrix eigenvalue problem, i.e., $\boldsymbol{\mathcal{H}}\overrightarrow{\psi_n} = \lambda_n\overrightarrow{\psi_n}$, $\omega_n^2 = -\lambda_n$, with $n$=1,2,3…. Here, $\overrightarrow{\psi_n}$ is the discrete sampling of $\psi_n(x_2)$. Dirichlet boundary conditions, i.e., $\psi_n(x_2)$=0, are applied to both ends of the 1D simulation system to represent the rigid domains ($\delta P_1(x_2,t) = 0$) on either side of the DW. The tridiagonal matrix $\boldsymbol{\mathcal{H}}$ is then diagonalized using standard symmetric eigensolvers, which directly yield the eigenvalues $\lambda_n$ and the corresponding normalized eigenvectors $\overrightarrow{\psi_n}$. The eigenfrequencies follow the relation $\omega_n^2 = -\lambda_n$, and the eigenvectors provide the spatial mode profiles of the domain-wall collective oscillations. As $\varepsilon_{11}$ increases, the term ($K_{11}^{\mathrm{Landau}}$+$K_{11}^{\mathrm{elas}}$) in Eq. (A1) becomes larger, leading to higher eigenfrequencies $\omega_n$.

As an example, we consider the case of $\varepsilon_{11}$=0.7% and extract the lowest three-order DW eigenmodes, as shown in Figs. 6(a-c). The first-order (*n*=1) eigenmode corresponds to a zero-energy sliding mode, with an eigenfunction proportional to the first-order spatial derivative of the equilibrium DW profile, i.e. $\psi_1(x_2) \propto \partial P_1^{\mathrm{eq}}(x_2)/\partial x_2$, which is an even function with respect to the DW center, i.e. $\psi_1(-x_2) = \psi_1(x_2)$. The eigenfrequency $\omega_1/2\pi$ is close to zero, reflecting the absence of a restoring force due to perfect translational invariance of the system. The second-order (*n*=2) eigenmode corresponds to a breathing mode, with an eigenfunction proportional to the second-order spatial derivative of the equilibrium DW profile, i.e., $\psi_3(x_2) \propto \partial^2 P_1^{\mathrm{eq}}(x_2)/\partial x_2^2$. This eigenfunction is zero (representing a node in a periodically oscillating wave) at the DW center, i.e. $\psi_2(x_2 = 0) = 0$, indicating that the location of the DW center is fixed. Moreover, the eigenfunction is an odd function with respect to the DW center, i.e., $\psi_2(-x_2) = -\psi_2(x_2)$, indicating that the polarization on the two sides of the DW oscillate along the opposite directions.

**Appendix B: Polarization profile/distribution reconstructed at a target frequency**

To reconstruct the line profiles in Figs. 3(g-i), we first select a line profile along the $x_2$ axis (i.e., the dashed line in Fig. 2(a)) to extract the raw simulation data $P_1(x_2, t)$. The Fourier analysis of this raw data gives a frequency spectrum for all the points along the selected line $\widetilde{P_1}(x_2, f)$. This intensity spectrum consists of multiple frequency peaks and includes all the dynamical components. To reconstruct the spatial profile of polarization oscillating at a target frequency, $\Delta P_1(x_2, t)$, we perform the inverse Fourier transform for the intensity spectrum $\widetilde{P_1}(x_2, f_0 - \Delta f < f < f_0 + \Delta f)$ at the target frequency peak $f_0$ with a selected bandwidth of 2$\Delta f$. The reconstructed line profiles are plotted as $P_1^{\mathrm{eq}}(x_2) + A\Delta P_1(x_2, t)$, where $P_1^{\mathrm{eq}}(x_2)$ is the initial equilibrium polarization profile at $t$=0, and the pre-factor $A$ is used to amplify the amplitude of polarization oscillation for visualization. We set $\Delta f$=0.05, 0.05, and 0.01 THz, $A$=3×10$^4$, 3×10$^4$, and 4×10$^2$ for Figs. 3(g,h,i), respectively.

Similarly, to reconstruct the 2D polarization distribution at the target frequency, we first select a 2D slice in the $x_1$ - $x_2$ plane and reconstruct both the $\Delta P_1(x_1, x_2, t)$ and $\Delta P_2(x_1, x_2, t)$ components using the inverse Fourier transform method described above. By plotting the reconstructed polarization distribution in the $x_1$-$x_2$ plane using vectors whose length indicates the amplitude of the polarization oscillation, we obtain the 2D distribution of polarization vector for the mode oscillating at the target frequency. The results for the unconventional DW sliding mode are shown in Fig. 4(a).

## Appendix C. Analytical model for the unconventional DW sliding mode

The uncompensated bulk polarization bound charge density produces a depolarization field, as shown in Figs. 4(a-c). The $x_1$ component of the depolarization field, $E_1^{\mathrm{d}}$, acts as a restoring force when the DW is displaced from its equilibrium position. To calculate the frequency of such bulk-polarization-charge induced DW sliding mode, $\omega_{\mathrm{DW}}$, we develop a harmonic oscillator model by analogy to that reported in [77], with an equation of motion written as,

$$\mu_{\mathrm{DW}}\frac{\partial^2 x(t)}{\partial t^2}+\gamma_{\mathrm{DW}}\frac{\partial x(t)}{\partial t}+k_{\mathrm{DW}}x(t)=0, \tag{C1}$$

where $x(t)$ denotes the DW displacement along the $x_2$ axis with respect to the equilibrium location (unit: m), $\mu_{\mathrm{DW}}$ is the DW mass density (unit: kg/m$^2$), $\gamma_{\mathrm{DW}}$ is the damping coefficient (unit: kg/m$^2$/s), and $k_{\mathrm{DW}}$ is the spring constant (unit: kg/m$^2$/s$^2$) which is associated with the areal potential energy $\Delta U$ (unit: J/m$^2$), written as, $\Delta U=\frac{1}{2}k_{\mathrm{DW}}x(t)^2$.Physically, $\Delta U$ originates from the local electrostatic energy penalty resulting from $E_1^{\mathrm{d}}$. Therefore, $\Delta U$ can be calculated by integrating the local electrostatic energy density, $-\frac{1}{2}E_1^{\mathrm{d}}(x_2)P_1(x_2)$ along the $x_2$ axis, i.e.,

$$\Delta U=\int_{-\frac{W}{2}+x}^{\frac{W}{2}}-\frac{1}{2}E_1^{\mathrm{d}}P_1(x_2-x)dx_2, \tag{C2}$$

where $W$ is the DW thickness. For simplification, we consider that the DW profile can be approximately described as $P_1(x_2)\approx -P_1^{0,\mathrm{eq}}$ when $x_2<-\frac{W}{2}$, $P_1(x_2)\approx\frac{P_1^{0,\mathrm{eq}}}{W}x_2$ when $-\frac{W}{2}\le x_2\le\frac{W}{2}$; and $P_1(x_2)\approx P_1^{0,\mathrm{eq}}$ when $x_2\ge\frac{W}{2}$. The depolarization field, $E_1^{\mathrm{d}}$, is given by $E_1^{\mathrm{d}}\approx-\frac{\delta P_1}{\kappa_0\kappa_{\mathrm{b}}}$, where $\delta P_1$ denotes the bulk polarization charge density along the $x_1$ axis and can be approximated based on the DW displacement, i.e., $\delta P_1\approx-\frac{2P_1^{0,\mathrm{eq}}}{W}x(t)$. Taken together, $\Delta U$ can be further written as,

$$\Delta U\approx\frac{{P_1^{0,\mathrm{eq}}}^2x(t)^2}{W^2\kappa_0\kappa_{\mathrm{b}}}\big(W-x(t)\big). \tag{C3}$$

Equation (C3) shows that both a positive and a negative DW displacement $x(t)$ increases the areal potential energy electrostatic energy by an amount of $\Delta U$. If assuming $W\gg x(t)$, one can further derive the spring constant $k_{\mathrm{DW}}$ as,

$$k_{\mathrm{DW}}=\frac{d^2\Delta U}{dx^2}\approx\frac{2{P_1^{0,\mathrm{eq}}}^2}{\kappa_0\kappa_{\mathrm{b}}W}, \tag{C4}$$

A knowledge of $k_{\mathrm{DW}}$ allows for calculating the frequency of DW sliding,

$$\omega_{\mathrm{DW}}=\sqrt{\frac{k_{\mathrm{DW}}}{\mu_{\mathrm{DW}}}}=\sqrt{\frac{2{P_1^{0,\mathrm{eq}}}^2}{\mu_{\mathrm{DW}}\kappa_0\kappa_{\mathrm{b}}W}}, \tag{C5}$$

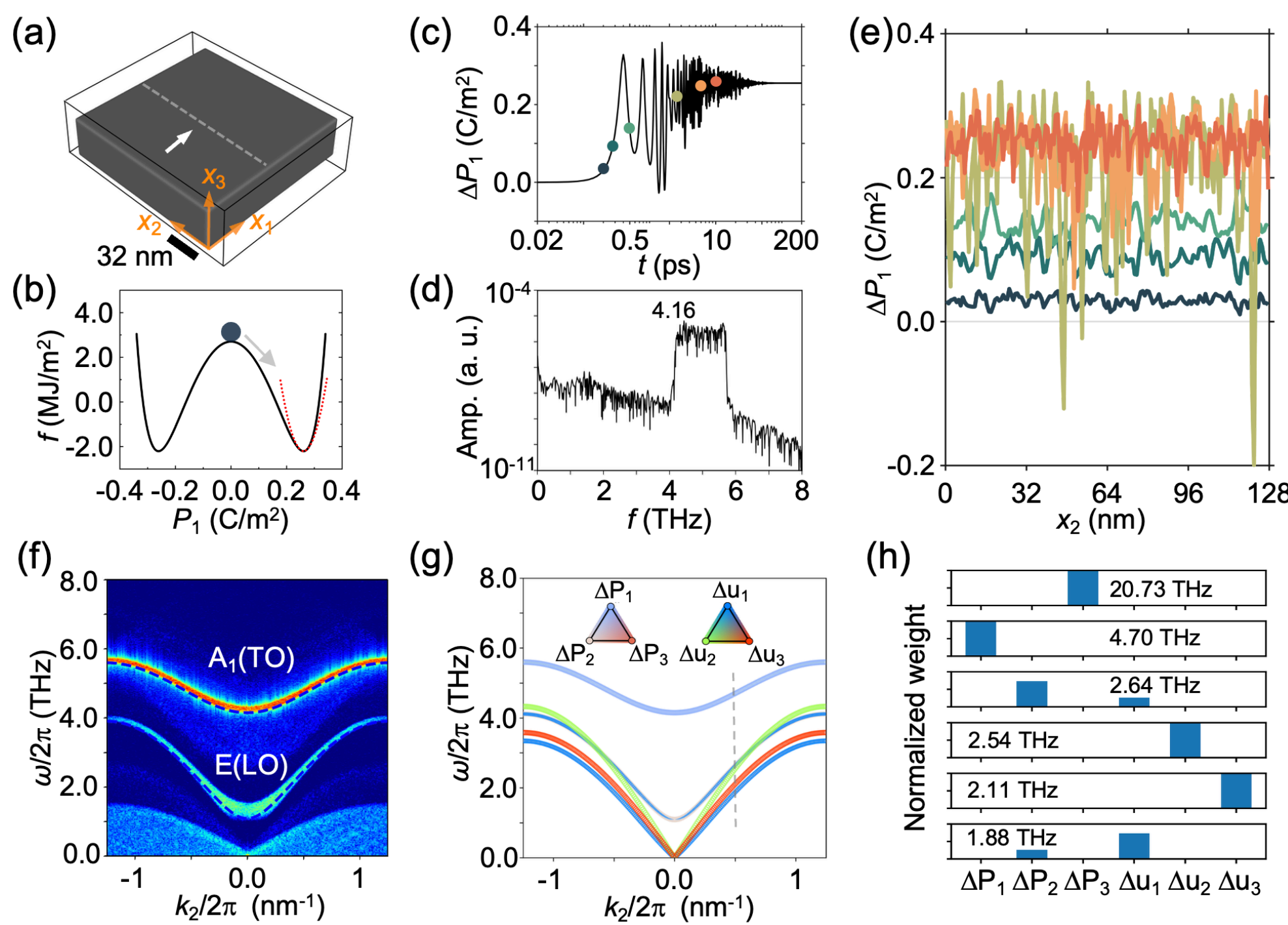

**Fig. 1. (a)** Equilibrium single-domain structure of a 32-nm-thick uniaxially strained $BaTiO_3$ membrane with $\varepsilon_{11}$=0.7%, obtained by dynamical phase-field simulations. The arrow indicates the direction of spontaneous polarization. The dashed line cuts through the middle plane at $x_3$=16 nm and $x_1$=64 nm. The cuboid with black frames represents the entire simulation system (see details in Sec. 2). (**b**) 1D free energy density of a single-domain $BaTiO_3$ membrane calculated by fixing the total strain tensor to those at the initial equilibrium condition ($P_i$=$P_i^{\text{eq}}$) at $\varepsilon_{11}$=0.70% and varying the eigenstrain $\varepsilon_i^0$ based on the $P_1$, with $P_2$=$P_3$=0. The red dashed line indicates the harmonic fitting of the free energy profile, which shows $|\Delta P_1|\leq 0.001$ C/m² within the harmonic regime. **(c)** Evolution of $\Delta P_1$ at the center of the simulation system where ($x_1$, $x_2$, $x_3$)=(64 nm, 64 nm, 16 nm) at 300 K. The system is in a paraelectric phase at $t$=0. **(d)** Frequency spectrum calculated by performing Fourier transform of the time-domain data in (**c**) for $t$=10-200 ps. **(e)** Spatial profiles of the $\Delta P_1$ along the dashed line in (**a**) at $t$=0.2 ps, 0.28 ps, 0.5 ps, 2.6 ps, 6.0 ps and 10.0 ps, corresponding to the dots in (**c**). **(f)** Dispersion relations of coherent polarization waves, calculated by performing 2D Fourier transform of the $\Delta P_1(x_2,t)$ along the dashed line in (**a**) for $t$=30-200 ps. The dashed curves indicate the analytically calculated dispersion of the $A_1$(TO) and E(LO) mode optical phonons (see Sec. 3.1). To mitigate spectral leakage, Hann window is applied to both the time-domain and space-domain data before the Fourier transform. **(g)** The analytically calculated dispersion curves for the coupled acoustic phonon ($\Delta u_i$) and optical phonon ($\Delta P_i$) branches, $i$=1,2,3, where the weighted projections of their dynamical components are denoted with color bar. The $\Delta P_3$ branch is above the 8 THz frequency range. **(h)** The normalized weighted projections of the $\Delta P_i$ and $\Delta u_i$ components for the six frequency modes at $k_2/2\pi$=0.5 nm⁻¹.

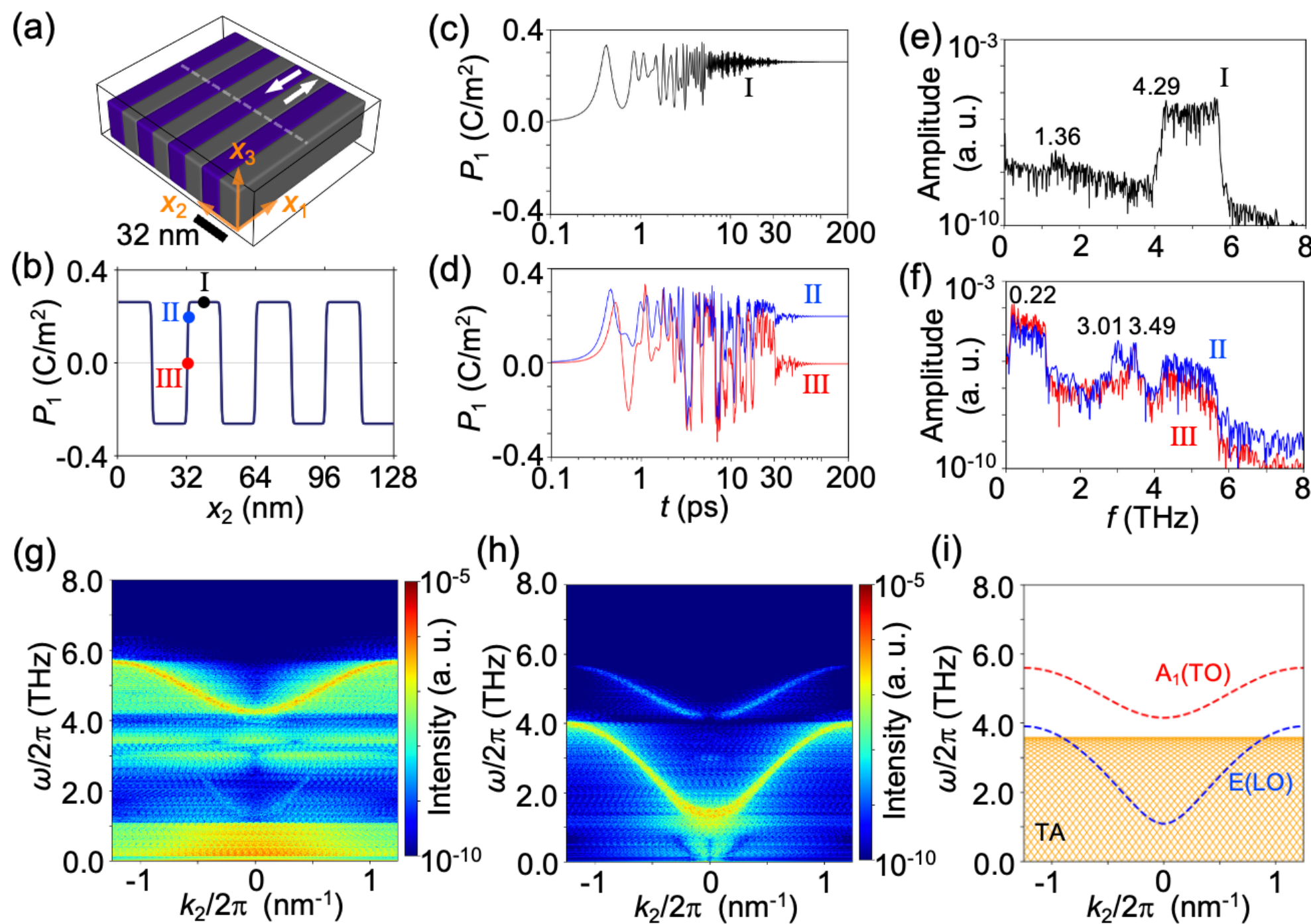

**Fig. 2. (a)** Equilibrium stripe domain structure for a 32-nm-thick uniaxially strained $BaTiO_3$ membrane with $\varepsilon_{11}$=0.7%, obtained by dynamical phase-field simulations. The arrows indicate the direction of local spontaneous polarization. The location of the dashed line is the same as that in Fig. 1(a). **(b)** The spatial profile of polarization $P_1$ along the dashed line in Fig. 2(a), where $P_2 \approx 0$, $P_3 \approx 0$. **(c)** Evolution of $\Delta P_1$ at the domain center, represented by the point I in (b). **(d)** Evolution of $\Delta P_1$ at the DW, represented by the blue (point II) and red (point III) points in (b). **(e,f)** Frequency spectra calculated by performing Fourier transform of the time-domain data in (c,d) for $t$=30-200 ps, respectively. Dispersion relations of coherent polarization waves, calculated by performing 2D Fourier transform of (**g**) $\Delta P_1(x_2,t)$ and (**h**) $\Delta P_2(x_2,t)$ along the dashed line in (**a**) for $t$=30-200 ps. To mitigate spectral leakage, Hann window is applied to both the time-domain and space-domain data before the Fourier transform. **(i)** Analytically calculated dispersion relations of the $A_1$(TO) and E(LO) mode optical phonons (see Sec. 3.1), and the band-folded transverse acoustic phonons (orange, see Sec. 3.3).

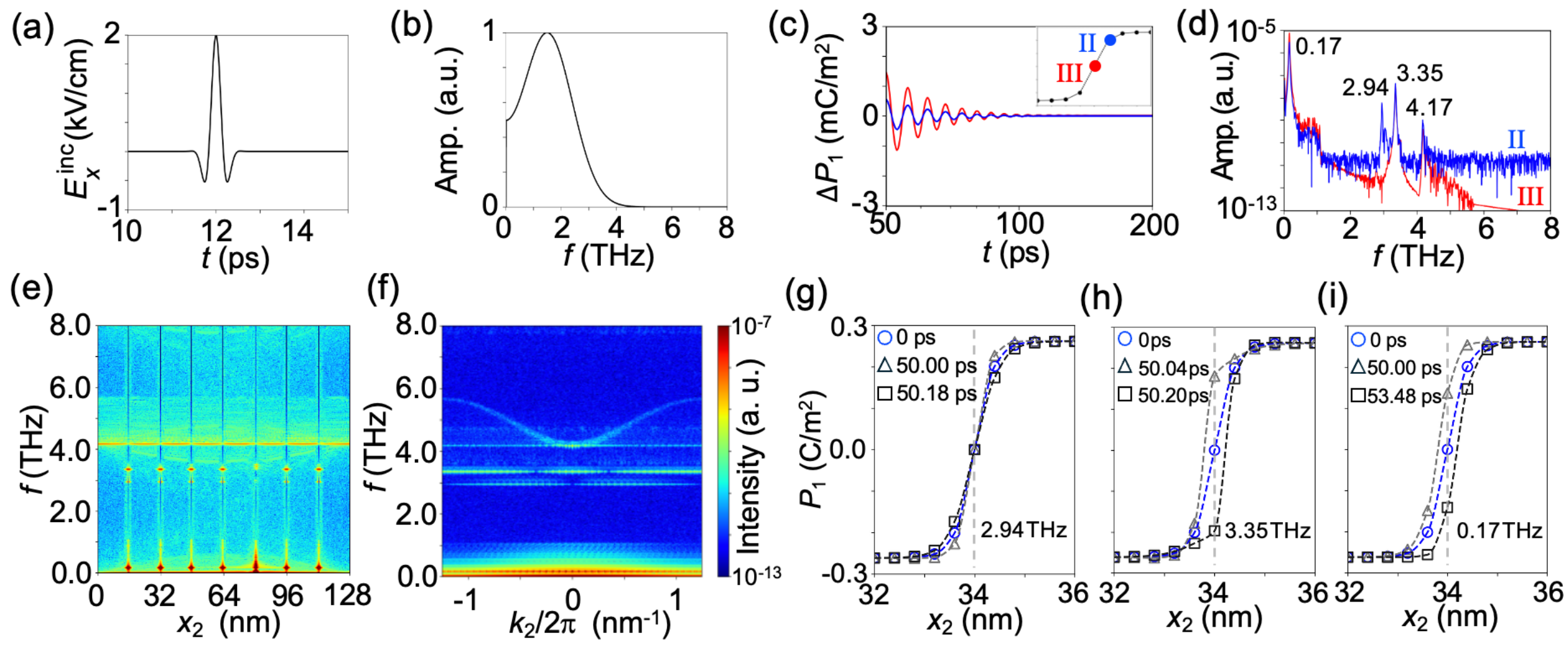

**Fig. 3. (a)** Temporal profile and **(b)** frequency spectrum of the broadband single-cycle electric-field pulse $E_x^{\mathrm{inc}}$ applied to the equilibrium ($t$=0) stripe domain structure in Fig. 2(a). (**c**) Evolution of $\Delta P_1$ at two single points at the DW, as indicated by red and blue color in the inset (the same as those in Fig. 2(b)), for $t$=50-200 ps. **(d)** Frequency spectra calculated by performing Fourier transform of the time-domain data in (**c**). **(e)** Spatially resolved frequency spectra obtained by time-domain Fourier transform of $\Delta P_1(x_2,t)$ along the same line shown in Fig. 2(a) for $t$=50-200 ps after the THz pulse excitation. **(f)** Dispersion relations of coherent polarization waves, calculated by performing 2D Fourier transform of the $\Delta P_1(x_2,t)$ for $t$=50-200 ps. To mitigate spectral leakage, Hann window is applied to both the time-domain and space-domain data before the Fourier transform. Line profiles of polarization at different time stages, which oscillate at approximately a single temporal frequency of (**g**) 2.94 THz, **(h)** 3.35 THz, and (**i**) 0.17 THz. The markers are the reconstructed polarization amplitude via the inverse Fourier transform (Appendix B), where the data points for "0 ps" indicate the equilibrium polarization profiles. The dashed curves are the approximate line profiles from eigenmode analysis (Appendix A).

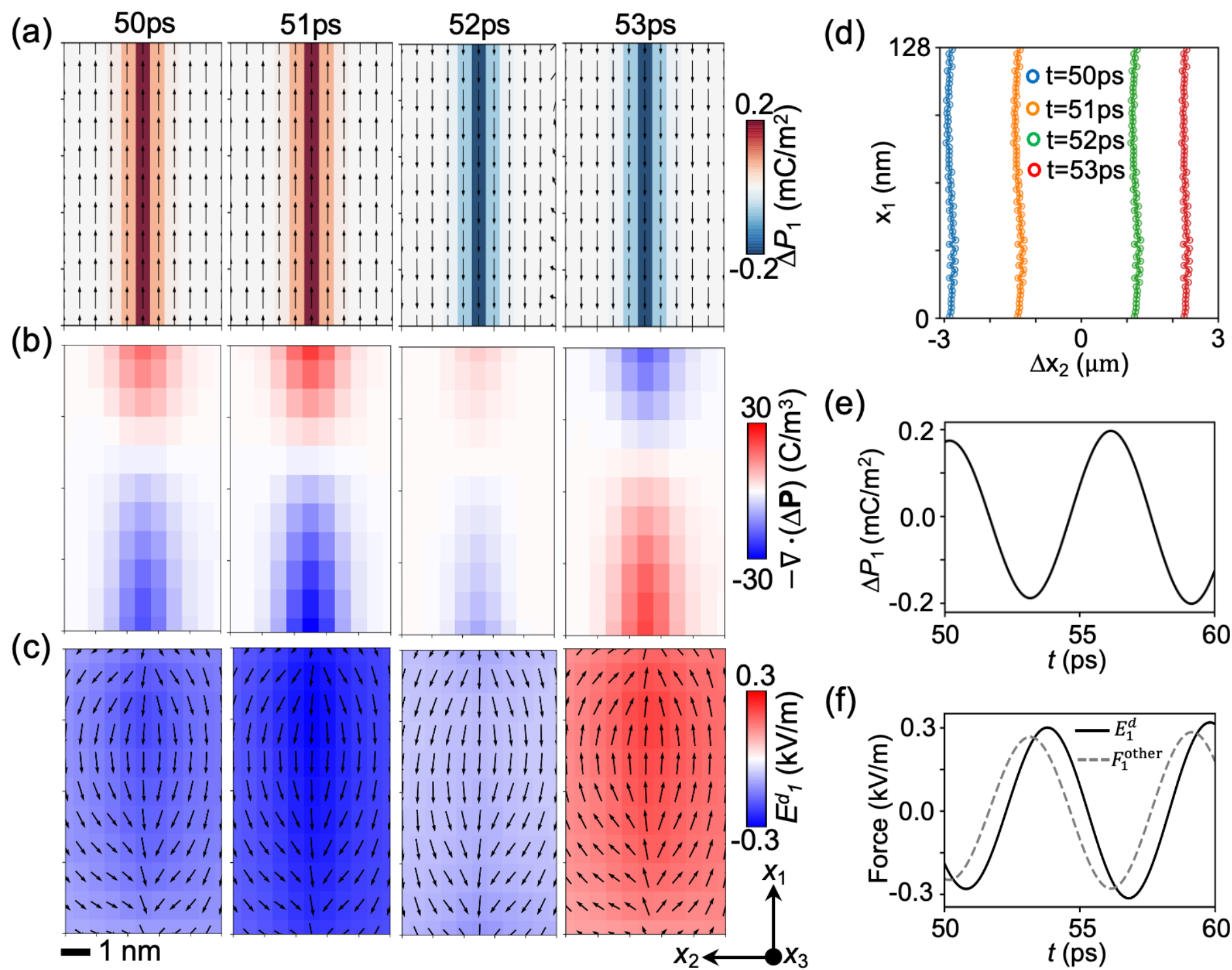

**Fig. 4.** (**a**) 2D distribution of polarization vector reconstructed via the inverse Fourier transform near the frequency of 0.17 THz from the data in Fig. 3 at different time stages (Appendix B). The 2D slices show the range of $x_1$=80-100 nm and $x_2$=32-36 nm at $x_3$=16 nm. $\Delta P_1(t)= P_1(t)-P_1^{\mathrm{eq}}$. (**b**) The distribution of the polarization bound charge $-\nabla\cdot(\Delta\mathbf{P})$ calculated from the reconstructed polarization vectors in (**a**). **(c)** The depolarization field distribution calculated from the reconstructed polarization vectors in (**a**). **(d)** The DW center offsets $\Delta x_2$ with respect to the equilibrium DW center (see the vertical gray dashed line in Fig. 3(i)) at different time stages. The location of the DW center is approximated as the $x_2$ coordinate where $P_1$=0. Evolution of (**e**) the reconstructed $\Delta P_1$ for a single point in the DW and **(f)** the corresponding restoring force components. $E_1^{\mathrm{d}}$ is the depolarization contribution, $F_1^{\mathrm{other}}$ is the sum of the other contributions including the Landau, elastic, and the gradient energy density (Sec. 3.1).

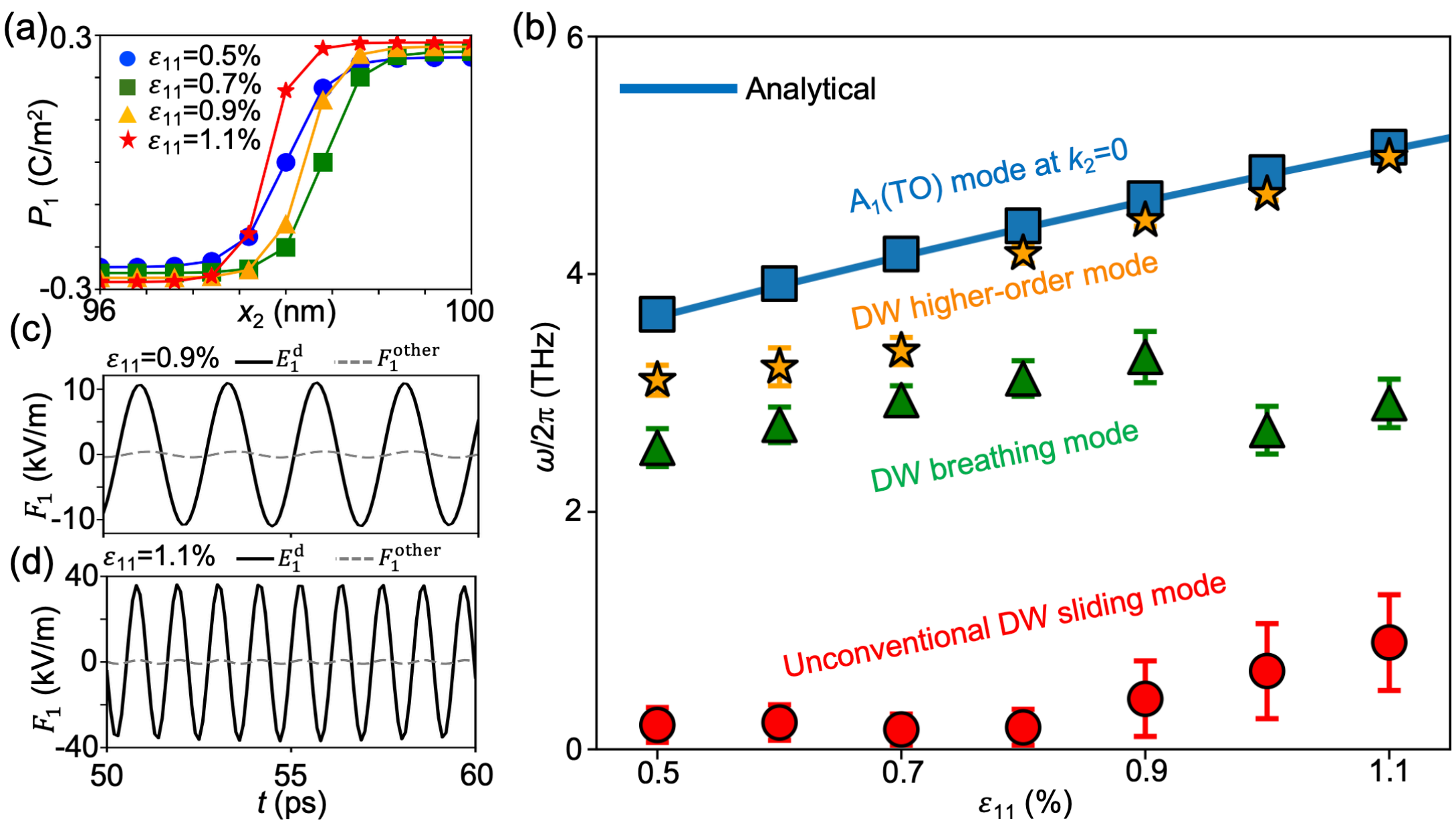

**Fig. 5. (a)** Equilibrium domain wall (DW) spatial profile for $\varepsilon_{11}$=0.5%, 0.7%, 0.9%, and 1.1%, obtained by dynamical phase-field simulations. **(b)** Strain modulation of the frequencies of the $A_1$(TO) mode (the ferroelectric soft mode), the DW breathing mode (the second-order eigenmode), the DW higher-order mode (the third-order eigenmode), and the unconventional DW sliding mode. Details of analytical calculation are in Sec. 3.1. The markers indicate the frequencies extracted from dynamical phase-field simulations. The error bars represent the full-width at half-maximum of a frequency peak. The $A_1$(TO) mode frequencies are taken based on the lowest frequency peaks that agree with the analytical calculation at $k_2$=0, thus have no error bar. The time evolution of the restoring force components along the $x_1$ direction at a single point in the DW for the unconventional DW sliding mode for **(c)** $\varepsilon_{11}$=0.9% and **(d)** $\varepsilon_{11}$=1.1%. $E_1^{\mathrm{d}}$ is the depolarization contribution, $F_1^{\mathrm{other}}$ is the sum of the other contributions including the Landau, elastic, and the gradient energy (see Sec. 3.1).

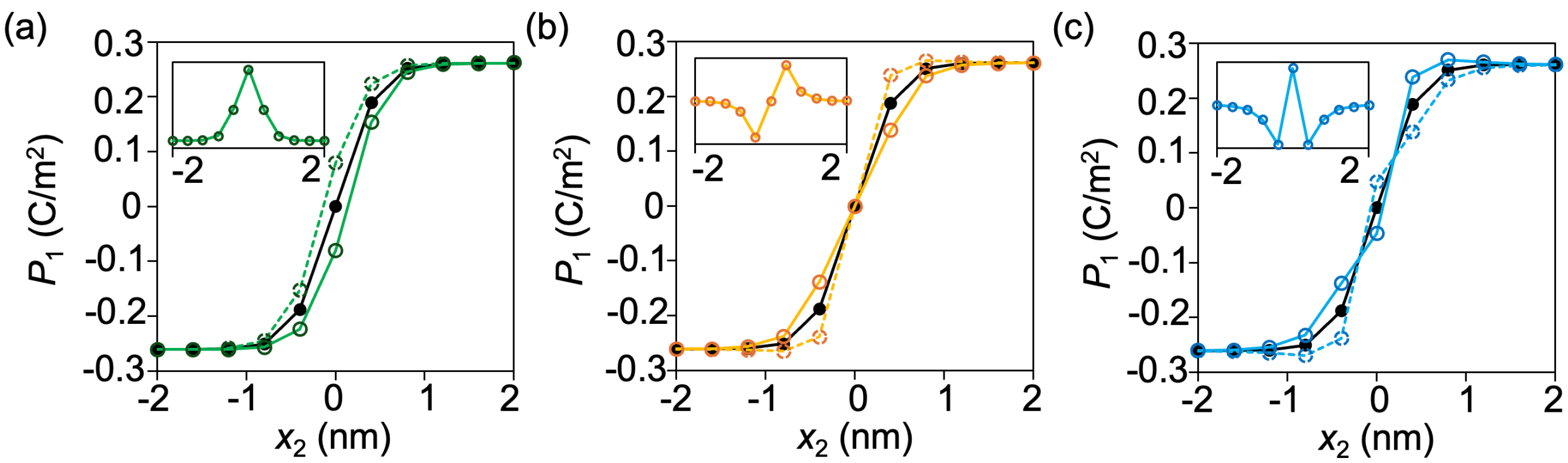

**Fig. 6.** Spatiotemporal evolution of the $P_1(x_2)$ associated with the (**a**) first-order ($n$=1), (**b**) second-order ($n$=2), and (**c**) third-order ($n$=3) eigenmode of a 1D Ising-type 180° DW, obtained from the 1D numerical eigenmode analysis (Appendix A). The DW center is located at $x_2$=0. The black curves with solid circles denote the equilibrium ($t$=0) DW profile $P_1^{\mathrm{eq}}(x_2)$, while the colored curves with hollow circles represent the polarization profile at different oscillation stages, i.e. $P_1(x_2) = P_1^{\mathrm{eq}}(x_2) \pm A\Delta P_1(x_2)$, where $\Delta P_1(x_2)$ denotes the polarization variation associated with the corresponding resonant mode shown in the inset, and $A$ is the pre-factor used to amplify the polarization oscillation amplitude for visualization. For clarity, the polarization variation $\Delta P_1(x_2)$ is normalized with its maximum amplitude is manually set to 1 C/m$^2$. The values of $A$ are 0.05, 0.05, and 0.08 for the modes in (**a**,**b**,**c**), respectively. The profiles of the eigenfunctions $\psi_n(x_2)$, $n$=1,2,3 are shown in the insets of (**a**,**b**,**c**), respectively.

**Supplementary Information**

Supplemental Figure 1: Domain wall modes with increased polarization damping $\gamma$.

Supplemental Figure 2: Distribution of the polarization bound charge components $-\partial \Delta P_1/\partial x_1$ and $-\partial \Delta P_2/\partial x_2$ calculated from the reconstructed polarization vectors in Fig. 5(a).

Supplemental Figure 3: Temperature-strain stability phase diagram of uniaxially strained $BaTiO_3$ membrane

Supplemental Figure 4: Domain wall modes in a uniaxially strained $BaTiO_3$ membrane with two domains and periodic boundary condition along the $x_2$ axis.

Supplemental Note 1. Full expressions of the matrices in Eqs. (11a-b) of the main text

Supplemental Note 2. Discussion on the temperature-strain stability phase diagram.

**Supplemental Figure 1.**

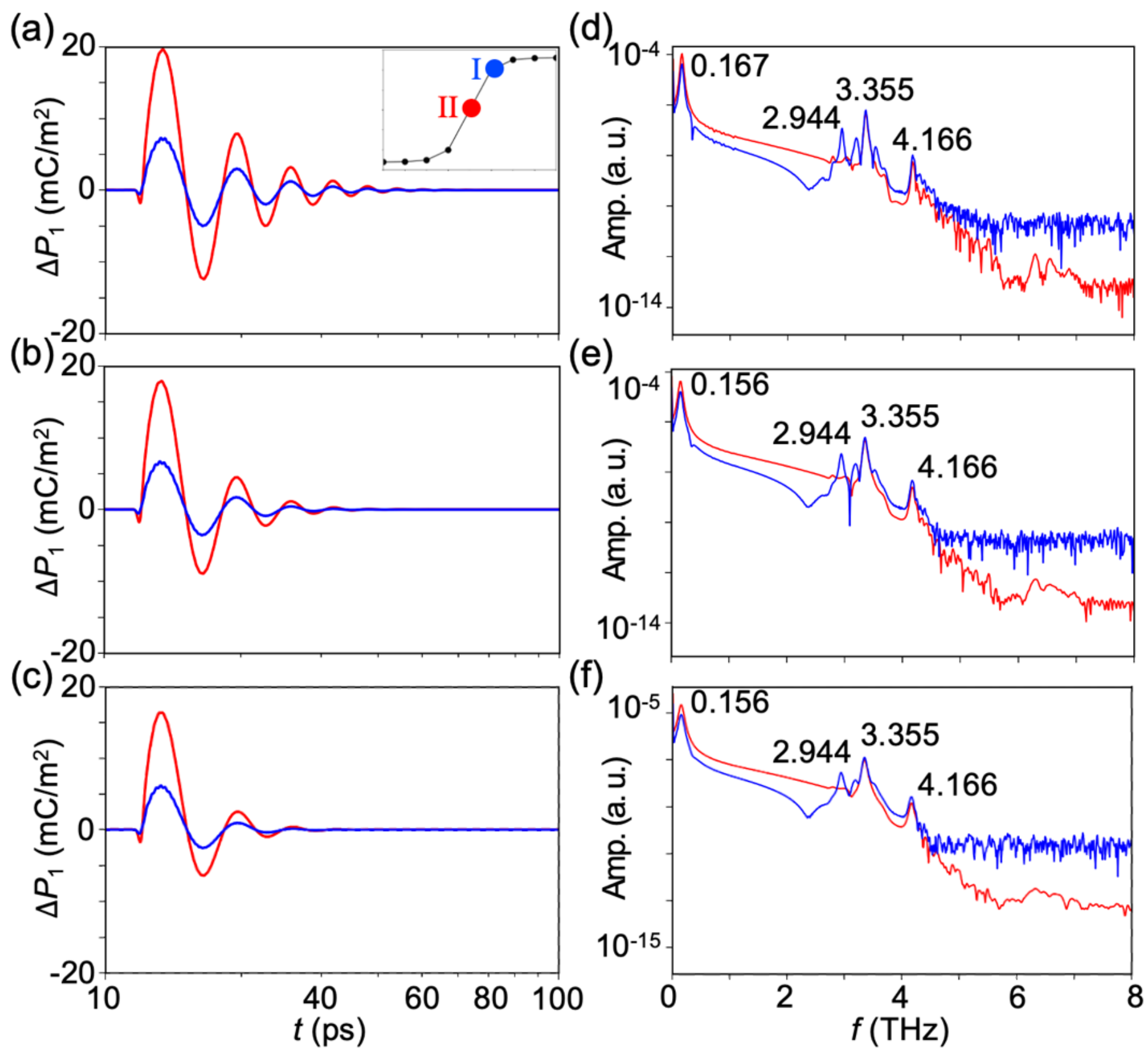

**Figure S1. (a,b,c)** Evolution of $\Delta P_1$ at two single points at the DW, as indicated by red and blue color in the inset, for $\gamma$=4×10$^{-7}$ Ω·m, $\gamma$=6×10$^{-7}$ Ω·m, and $\gamma$=8×10$^{-7}$ Ω·m, respectively. **(d,e,f)** Frequency spectra calculated by performing Fourier transform of the time-domain data in (a,b,c), for $t$=10-100 ps. To mitigate spectral leakage, Hann window is applied to the time-domain data before the Fourier transform.

## Supplemental Figure 2.

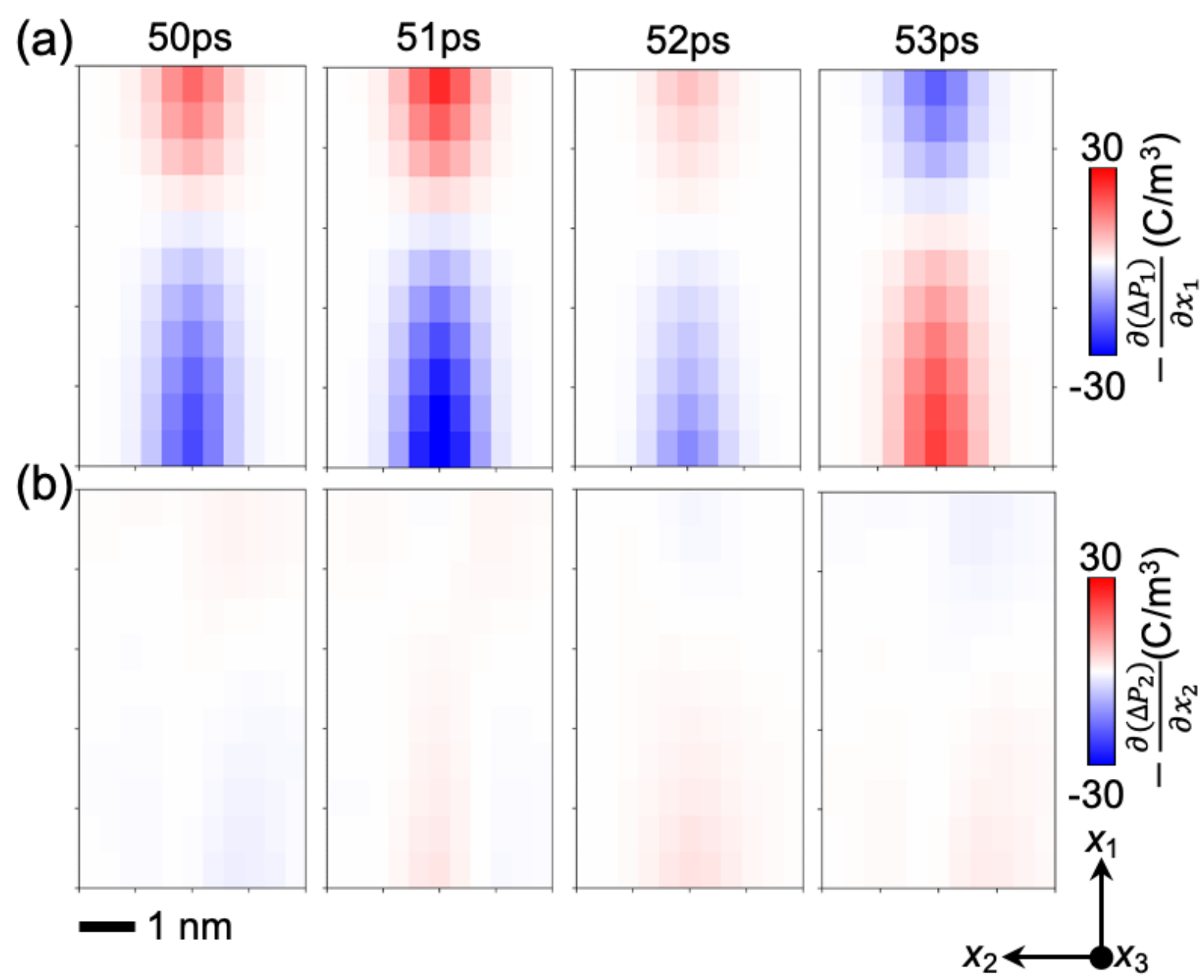

**Figure S2. (a,b)** The distribution of the polarization bound charge components $-\partial\Delta P_1/\partial x_1$ and $-\partial\Delta P_2/\partial x_2$ calculated from the reconstructed polarization vectors in Fig. 5(a) in the main text.

## Supplemental Figure 3. Temperature-strain stability phase diagram of uniaxially strained $BaTiO_3$ membrane

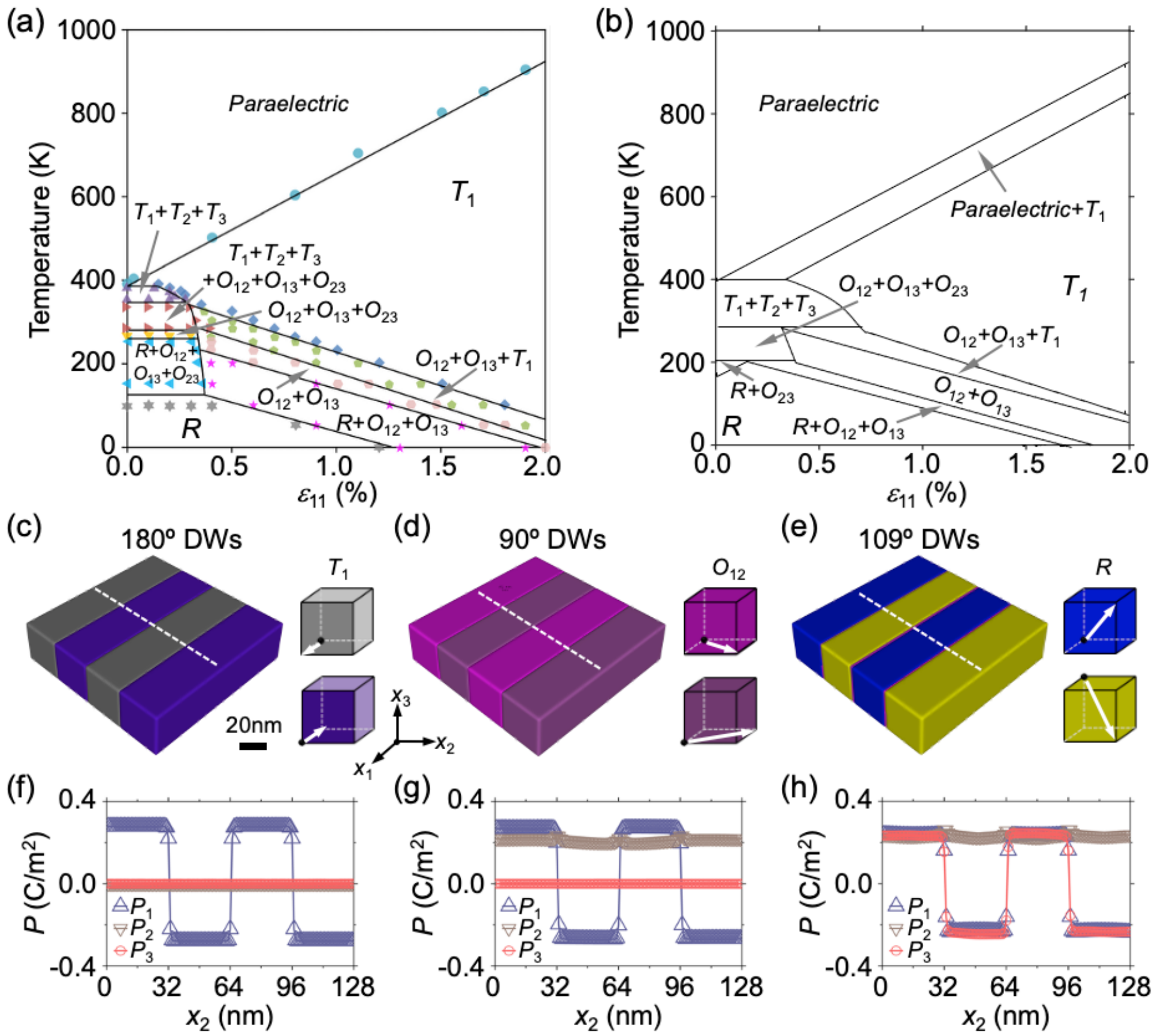

**Figure S3.** The temperature-uniaxial strain stability phase diagram of multidomain structures of the BTO membrane from **(a)** dynamical phase-field simulations and (**b**) analytical calculations. In dynamical phase-field simulations, the total strain $\varepsilon_{11}$ varies between 0% and 2% with a stepping of 0.1%, and the temperature vary from 0 K to 1000 K with a stepping of 10 K. **(c,d,e)** Typical striped domain structures of the BTO membrane obtained under uniaxial strain of 1.0% at 300 K (c), 0.7% at 200 K (d), and 0.2% at 100 K (e), from the phase diagram in (a). The initial polarization distributions were defined as follows: for 180º DW, opposite $\pm P_1$ domains with $P_2=P_3=0$; for 90º DW, alternating $P_1$ while keeping $P_2$ uniform ($P_3$=0); for 109º DW, opposite $P_1$ and $P_3$ while $P_2$ remained uniform between domains. In all cases, the initial $|P_i|$ were assigned within 0.01 C/m$^2$ to provide small perturbations for domain evolution. (**f**,**g**,**h**) The spatial profiles of the polarization components $P_1$, $P_2$, and $P_3$ along the dashed lines in (c-e), corresponding to the 180º, 90º, and 109º DWs, respectively.

## Supplemental Figure 4. Domain wall modes in a uniaxially strained $BaTiO_3$ membrane with two domains and periodic boundary condition along the $x_2$ axis

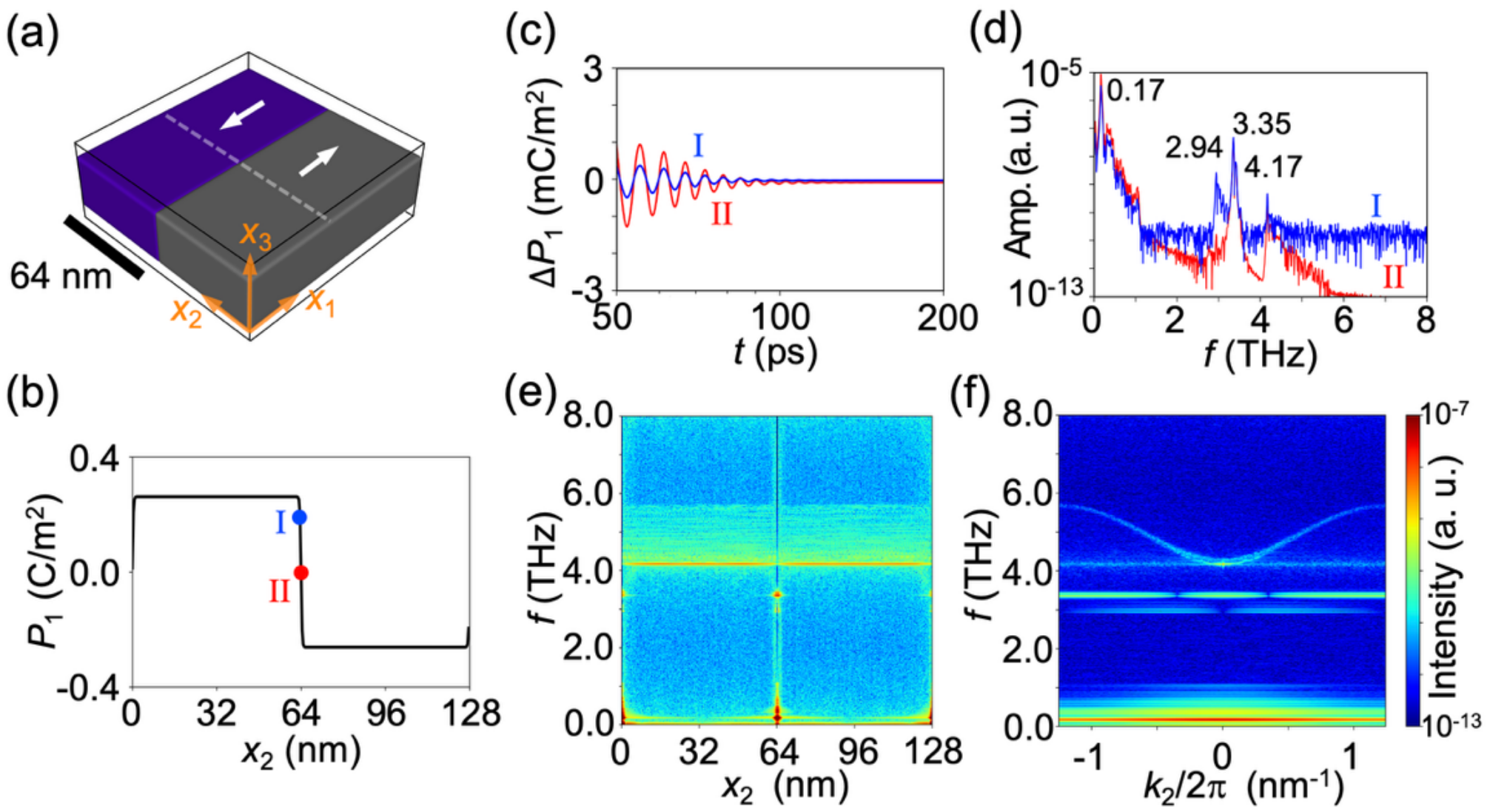

**Figure S4. (a)** Equilibrium stripe domain structure for a 32-nm-thick uniaxial strained $BaTiO_3$ membrane with $\varepsilon_{11}$=0.7%, obtained by dynamical phase-field simulations. Periodic boundary condition is now applied along the $x_2$ axis by removing the air phase. The arrows indicate the direction of local spontaneous polarization. The location of the dashed line is the same as that in the Fig. 2(a) of the main text. **(c)** The spatial profile of polarization $P_1$ along the dashed line in Fig. 1(a), where $P_2 \approx 0$, $P_3 \approx 0$. Evolution of $\Delta P_1$ at two single points at the DW, as indicated by red and blue color in the inset, for $t$=50-200 ps. **(d)** Frequency spectra calculated by performing Fourier transform of the time-domain data in (**c**). **(e)** Spatially resolved frequency spectra obtained by time-domain Fourier transform of $\Delta P_1(x_2,t)$ along the same line shown in Fig. 2(a) for $t$=50-200 ps after the THz pulse excitation. **(f)** Dispersion relation of coherent polarization wave, calculated by performing 2D Fourier transform of the $\Delta P_1(x_2,t)$ for $t$=50-200 ps.

**Supplemental Note 1. Full expressions of the matrices in Eqs. (12a-b) of the main text**

$$\mathbf{R}=\begin{pmatrix}\frac{\partial}{\partial x_1} & 0 & 0\\ 0 & \frac{\partial}{\partial x_2} & 0\\ 0 & 0 & \frac{\partial}{\partial x_3}\\ 0 & \frac{1}{2}\frac{\partial}{\partial x_3} & \frac{1}{2}\frac{\partial}{\partial x_2}\\ \frac{1}{2}\frac{\partial}{\partial x_3} & 0 & \frac{1}{2}\frac{\partial}{\partial x_1}\\ \frac{1}{2}\frac{\partial}{\partial x_2} & \frac{1}{2}\frac{\partial}{\partial x_1} & 0\end{pmatrix},$$

$$\widetilde{\mathbf{R}}=\mathrm{diag}(1,1,1,2,2,2)$$

$$\widehat{\mathbf{R}}=\begin{pmatrix}\frac{\partial}{\partial x_1} & 0 & 0 & 0 & \frac{\partial}{\partial x_3} & \frac{\partial}{\partial x_2}\\ 0 & \frac{\partial}{\partial x_2} & 0 & \frac{\partial}{\partial x_3} & 0 & \frac{\partial}{\partial x_1}\\ 0 & 0 & \frac{\partial}{\partial x_3} & \frac{\partial}{\partial x_2} & \frac{\partial}{\partial x_1} & 0\end{pmatrix},$$

$$\widehat{\mathbf{P}}^{\mathrm{eq}}=\begin{pmatrix}2P_1^{\mathrm{eq}} & 0 & 0\\ 0 & 2P_2^{\mathrm{eq}} & 0\\ 0 & 0 & 2P_3^{\mathrm{eq}}\\ 0 & P_3^{\mathrm{eq}} & P_2^{\mathrm{eq}}\\ P_3^{\mathrm{eq}} & 0 & P_1^{\mathrm{eq}}\\ P_2^{\mathrm{eq}} & P_1^{\mathrm{eq}} & 0\end{pmatrix},$$

$$\mathbf{R}(\mathbf{k})=\mathbf{i}\sqrt{2\big(1-\cos(k_2a)\big)/a^2}\begin{pmatrix}0 & 0 & 0\\ 0 & 1 & 0\\ 0 & 0 & 0\\ 0 & 0 & 1/2\\ 0 & 0 & 0\\ 1/2 & 0 & 0\end{pmatrix},$$

$$\widehat{\mathbf{R}}(\mathbf{k})=\mathbf{i}\sqrt{2\big(1-\cos(k_2a)\big)/a^2}\begin{pmatrix}0 & 0 & 0 & 0 & 0 & 1\\ 0 & 1 & 0 & 0 & 0 & 0\\ 0 & 0 & 0 & 1 & 0 & 0\end{pmatrix}.$$

**Supplemental Note 2. Discussion on the temperature-strain stability phase diagram**

Figure S1(a) shows the temperature-uniaxial strain stability diagram of multiple polar phases (domain states) in a 32-nm-thick $BaTiO_3$ membrane obtained by dynamical phase-field simulations. Since the goal is to simulate equilibrium ferroelectric domain structure, we used a larger polarization damping coefficient $\gamma=2\times10^{-4}$ $\Omega\cdot$m and a larger time step $\Delta t=5\times10^{-15}$ s to ensure a faster evolution to thermodynamic equilibrium. A larger gradient energy $G_0=1.0\times10^{-10}$ $C^{-2}\cdot m^4\cdot N$ is also used. In this case, the domain wall width would be larger such that a larger simulation cell size (1 nm) can be used to resolve the domain wall. Specifically, in the simulations in Fig. S1, discretized cells of $N_1\Delta x_1 \times N_2\Delta x_2 \times N_3\Delta x_3$, with a cell size of $\Delta x_1=\Delta x_2=\Delta x_3$ = 1 nm and ($N_1$, $N_2$, $N_3$=128, 128, 32), are used to describe the ferroelectric membrane. The other simulation set-up is kept the same as the simulation of coupled polarization-strain dynamics in the main text.

As shown in Fig. S1(a), the membrane undergoes a cubic (paraelectric)-tetragonal($T$)-orthorhombic($O$)-rhombohedral ($R$) transition as temperature decreases, similarly to the bulk $BaTiO_3$ single crystals. At large $\varepsilon_{11}$, the $T_1$ phase (with a spontaneous polarization aligning along the $x_1$ axis) is stabilized. Figures 6(c-e) show the representative domain structures of $T$-, $O$-, and $R$- dominant phase under different temperatures and strain combinations with 180º, 90º, and 109º DW, and their corresponding line profiles of polarization are shown in Fig. 6(f-h), respectively. Figure 6(b) shows the phase diagram predicted by extending the thermodynamic theory of strain phase separation reported in [R1] to incorporate the mechanical boundary condition of a uniaxially strained membrane (see details in the section "*Analytical calculation of the temperature-uniaxial strain stability diagram for membranes*" below). This analytically calculated diagram reproduces the key features observed in phase-field simulations, including the locations of phase stability regions and the temperature-dependent slopes of phase boundaries. However, notable discrepancies arise in the stability of multiphase regions. Because the thermodynamic theory neglects the energetic cost of spatial inhomogeneities, such as domain walls and the effects of inhomogeneous **P** on the local total strain and depolarization field, it tends to overestimate the stability window of multidomain configurations. For instance, the strain phase-separation model predicts the $(T_1+T_2+T_3)/T_1$ phase boundary near 0.65% strain at 300 K, whereas phase-field simulations place the boundary closer to 0.4% strain. In some cases, the inclusion of spatial inhomogeneity energy in the phase-field model eliminates entire phase regions predicted by the thermodynamic approach. One such example is the paraelectric + $T_1$ region, which appears in the strain phase-separation calculations but is absent in the phase-field results. Such agreements and the explainable discrepancies together corroborate the thermodynamical consistency of the equilibrium ferroelectric domain structures predicted by dynamical phase-field simulations.

*Analytical calculation of the temperature-uniaxial strain stability diagram for membranes*

We start from the Gibbs free energy density function under constant temperature and stress conditions,

$$g\left(T, P_i, \sigma_{ij}\right) = \alpha_1(P_1^2 + P_2^2 + P_3^2) + \alpha_{11}(P_1^4 + P_2^4 + P_3^4) + \alpha_{12}(P_1^2P_2^2 + P_2^2P_3^2 + P_1^2P_3^2)$$
$$+\alpha_{111}(P_1^6 + P_2^6 + P_3^6) + \alpha_{112}(P_1^2(P_2^4 + P_3^4) + P_2^2(P_1^4 + P_3^4) + P_3^2(P_2^4 + P_3^4)) +$$
$$\alpha_{123}(P_1^2P_2^2P_3^2) + \alpha_{1111}(P_1^8 + P_2^8 + P_3^8) + \alpha_{1112}[P_1^6(P_2^2 + P_3^2) + P_2^6(P_1^2 + P_3^2) +$$
$$P_3^6(P_1^2 + P_2^2)] + \alpha_{1122}(P_1^4P_2^4 + P_2^4P_3^4 + P_1^4P_3^4) + \alpha_{1123}(P_1^4P_2^2P_3^2 + P_2^4P_1^2P_3^2 + P_3^4P_2^2P_1^2)$$
$$-\frac{1}{2}S_{11}(\sigma_{11}^2 + \sigma_{22}^2 + \sigma_{33}^2) - S_{12}(\sigma_{11}\sigma_{22} + \sigma_{22}\sigma_{33} + \sigma_{33}\sigma_{11}) - \frac{1}{2}S_{44}(\sigma_{12}^2 + \sigma_{23}^2 + \sigma_{31}^2)$$
$$-Q_{11}(\sigma_{11}P_1^2 + \sigma_{22}P_2^2 + \sigma_3P_3^2) - Q_{12}(\sigma_{11}(P_2^2 + P_3^2) + \sigma_{22}(P_1^2 + P_3^2) + \sigma_{33}(P_1^2 + P_2^2))$$
$$-2Q_{44}(\sigma_{23}P_2P_3 + \sigma_{13}P_1P_3 + \sigma_{12}P_1P_2), \tag{S1}$$

where $S_{11}$, $S_{12}$, and $S_{44}$ are elastic compliance coefficients. Next, we perform the Legendre transform to convert to a modified Helmholtz free energy, going from constant stress to a uniaxially strained conditions,

$$\tilde{f}\left(T, P_i, \varepsilon_{11}, \sigma_{ij_{(ij\neq 11)}}\right) = f\left(T, P_i, \sigma_{ij}\right) + \varepsilon_{11}\sigma_{11}, \tag{S2}$$

where the total strain $\varepsilon_{11}$ can be derived via the relation,

$$\varepsilon_{11} = -\left(\frac{\partial f}{\partial \sigma_{11}}\right)_{T, P_i, \sigma_{ij(ij\neq 11)}} = Q_{11}P_1^2 + Q_{12}(P_2^2 + P_3^2) + S_{11}\sigma_{11} + S_{12}(\sigma_{22} + \sigma_{33}). \tag{S3}$$

Re-writing Eq. (S3) leads to,

$$\sigma_{11} = \frac{\varepsilon_{11} - Q_{11}P_1^2 - Q_{12}(P_2^2 + P_3^2) - S_{12}(\sigma_{22} + \sigma_{33})}{S_{11}}. \tag{S4}$$

Substituting Eq. (S4) into Eq. (S2), and setting $\sigma_{ij_{(ij\neq 11)}} = 0$ (or equivalently, $\sigma_{i2} = \sigma_{i3} = 0$, see also Eqs. (3a-c) in the main text) based on the mechanical boundary condition of the uniaxially strained membrane, the modified Helmholtz free energy can be expanded into,

$$\tilde{f}\left(T, P_i, \varepsilon_{11}, \sigma_{ij_{(ij\neq 11)}} = 0\right)$$
$$= a_1(P_1^2 + P_2^2 + P_3^2) + a_{11}(P_1^4 + P_2^4 + P_3^4) + a_{12}(P_1^2P_2^2 + P_2^2P_3^2 + P_1^2P_3^2)$$
$$+a_{111}(P_1^6 + P_2^6 + P_3^6) + a_{112}(P_1^2(P_2^4 + P_3^4) + P_2^2(P_1^4 + P_3^4) + P_3^2(P_2^4 + P_3^4)) +$$
$$a_{123}(P_1^2P_2^2P_3^2) + \frac{1}{2S_{11}}(Q_{11}P_1^2 + Q_{12}(P_2^2 + P_3^2) - \varepsilon_{11})^2, \tag{S5}$$

which simplifies to

$$\tilde{f}\left(T, P_i, \varepsilon_{11}, \sigma_{ij_{(ij\neq 11)}} = 0\right)$$
$$= a_1(P_1^2 + P_2^2 + P_3^2) + a_{11}(P_1^4 + P_2^4 + P_3^4) + a_{12}(P_1^2P_2^2 + P_2^2P_3^2 + P_1^2P_3^2)$$
$$+a_{111}(P_1^6 + P_2^6 + P_3^6) + a_{112}(P_1^2(P_2^4 + P_3^4) + P_2^2(P_1^4 + P_3^4) + P_3^2(P_2^4 + P_3^4)) +$$
$$a_{123}(P_1^2P_2^2P_3^2) + \frac{1}{2S_{11}}(\varepsilon_{11}^o - \varepsilon_{11})^2. \tag{S6}$$

To compute the $T$-$\varepsilon_{11}$ stability diagram. we minimize the modified Helmholtz free energy with respect to polarization over the range of possible temperature ($T$) and strain ($\varepsilon_{11}$) conditions to find the homogenous modified Helmholtz free energy of the system. Next, from the thermodynamic theory of strain phase separation [R1], we assume that the interfaces between different ferroelectric phases are incoherent, and they possess a negligible interfacial energy. Therefore, we may define the total modified Helmholtz Free Energy density as

$$\tilde{f}^{mix} = \omega^{\alpha} \tilde{f}^{\alpha} + \omega^{\beta} \tilde{f}^{\beta}, \tag{S7}$$

where thermodynamic equilibrium is achieved when $\tilde{f}^{mix}$ is minimized subject to the mechanical constraints that require,

$$\varepsilon_{11} = \omega^{\alpha} \varepsilon_{11}^{\alpha} + \omega^{\beta} \varepsilon_{11}^{\beta}, \tag{S8}$$

and the requirement that the total volume fraction must equal one, $\omega^{\alpha} + \omega^{\beta} = 1$, which may be found using the common tangent construction approach.

In ferroelectric systems, multiple ferroelectric phases can coexist with identical free-energy profiles as a function of strain, allowing for the existence of more ferroelectric variants, than one may initial expect from Gibbs phase rule. For example, for a tetragonal domain there are six possible ferroelectric variants, $(\pm P_1, 0, 0)$, $(0, \pm P_2, 0)$ and $(0, 0, \pm P_3)$, which can be referred to as $\pm T_1$, $\pm T_2$, and $\pm T_3$. Under a uniaxial strain, these six possible variants possess, only two unique energy surfaces: one for the $\pm a_1$ and one for the $\pm a_2$, and $\pm a_3$, i.e.,

$$f^{\alpha}(\varepsilon_{11}) = f^{+T_1}(\varepsilon_{11}) = f^{-T_1}(\varepsilon_{11}), \tag{S9a}$$

$$f^{\beta}(\varepsilon_{11}) = f^{+T_2}(\varepsilon_{11}) = f^{-T_2}(\varepsilon_{11}) = f^{+T_3}(\varepsilon_{11}) = f^{-T_3}(\varepsilon_{11}), \tag{S9b}$$

Consequently, the common tangent construction can calculate the volume fraction of each distinct branch (i.e, $\alpha$ and $\beta$), but not the individual volume fractions of each degenerate variant within a given branch.

## Supplemental Reference